\documentclass[12pt,onecolumn,draftcls]{IEEEtran}
\usepackage{amsfonts,amsmath,graphicx,epsfig}
\usepackage{amssymb,amsxtra,color}
\usepackage[ansinew]{inputenc}

\newtheorem{cor}{Corollary}

\newtheorem{remark}{Remark}
\newcommand{\prob}{{\cal P}}
        \newtheorem{definition}{Definition}%
        \newtheorem{theorem}{Theorem}%
        \newtheorem{lemma}{Lemma}%

\DeclareOldFontCommand{\rm}{\normalfont\rmfamily}{\mathrm}
\DeclareOldFontCommand{\sf}{\normalfont\sffamily}{\mathsf}
\DeclareOldFontCommand{\tt}{\normalfont\ttfamily}{\mathtt}
\DeclareOldFontCommand{\bf}{\normalfont\bfseries}{\mathbf}
\DeclareOldFontCommand{\it}{\normalfont\itshape}{\mathit}
\DeclareOldFontCommand{\sl}{\normalfont\slshape}{\@nomath\sl}
\DeclareOldFontCommand{\sc}{\normalfont\scshape}{\@nomath\sc}

\newcommand{\comment}[1]{}

\newcommand{\bml}[1]{\begin{multline}\label{#1}}
\newcommand{\eml}{\end{multline}}
\newcommand{\beq}[1]{\begin{equation}\label{#1}}
\newcommand{\eeq}{\end{equation}}

\newcommand{\beann}{\begin{eqnarray*}}
\newcommand{\eeann}{\end{eqnarray*}}
\newcommand{\bea}[1]{\begin{eqnarray}\label{#1}}
\newcommand{\eea}{\end{eqnarray}}
\newcommand{\bmp}{\begin{minipage}}
\newcommand{\emp}{\end{minipage}}

\newcommand{\definref}[1]{{\itshape Definition~\ref{#1}}}
\newcommand{\theorref}[1]{{\itshape Theorem~\ref{#1}}}
\newcommand{\lemmaref}[1]{{\itshape Lemma~\ref{#1}}}
\newcommand{\corolref}[1]{{\itshape Corollary~\ref{#1}}}

\newcommand{\secref}[1]{Section~\ref{#1}}

\newcommand{\fig}[1]{{Fig.~\ref{#1}}}

\newcommand{\expectation}[1]{E\{#1\}}

\def\SET0N {I\hspace{-0.8ex}N_0}

{%
}%

{%
}%
\newsavebox{\Citname}

\newcommand{\ignore}[1]{}

\newenvironment{proofmain}
  {{\it Proof of \theorref{th:main}:} \hspace{3mm}}
  {\par\vspace*{3mm}}

\newenvironment{proofdrn}
  {{\it Proof of \theorref{th:drn}:} \hspace{3mm}}
  {\par\vspace*{3mm}}

\newcommand{\proofend}{\hfill$\blacksquare$}

\begin{document}

\title{A Max-Flow Min-Cut Theorem with Applications in Small Worlds and \\
Dual Radio Networks}
\author{Rui A. Costa \hspace{1cm} Jo\~ao Barros\vspace{-1cm}
\thanks{Rui A. Costa is with the Instituto de Telecomunica\c{c}\~oes and the Departamento de Ciência dos Computadores da Faculdade de Ciências da Universidade do Porto, Porto, Portugal; URL: {\tt http://www.dcc.fc.up.pt/\~{ }ruicosta/}.
João Barros is with the Instituto de Telecomunica\c{c}\~oes and the Departamento de Engenharia Electrotécnica e de Computadores da Faculdade de Engenharia da Universidade do Porto, Porto, Portugal; URL: {\tt http://paginas.fe.up.pt/\~{ }jbarros/}.
This work was supported by the Funda\c{c}\~ao para a Ci\^encia e Tecnologia (Portuguese Foundation for Science and Technology)
under grants SFRH-BD-27273-2006 and POSC/EIA/62199/2004.
Parts of this work have been presented at ITW 2006~\cite{Costa-Barros:06a}, NetCod 2006~\cite{Costa-Barros:06b}, and SpaSWiN 2007~\cite{costa2007drn}.  }}

\maketitle

\begin{abstract}
Intrigued by the capacity of random networks, we start by proving a max-flow min-cut theorem that
is applicable to any random graph obeying a suitably defined independence-in-cut
property. We then show that this property is satisfied by relevant
classes, including small world topologies, which are pervasive
in both man-made and natural networks, and wireless networks of dual devices, which exploit
multiple radio interfaces to enhance the connectivity of the network. In both cases, we are able to
apply our theorem and derive max-flow min-cut bounds for network information flow.
\end{abstract}

\section{Introduction}
\label{sec:intro}

In the quest for the fundamental limits of communication networks, whose topology is typically described by graphs, the connection between the maximum information flow and the minimum cut of the network plays a singular and prominent role. In the case where the network has one or more independent sources of information but only one sink, it is known that the transmitted information behaves like {\it water in pipes} and the capacity can be obtained by classical network flow methods. Specifically, the capacity of this network will then follow from the well-known Ford-Fulkerson {\it max-flow min-cut} theorem~\cite{ford:62}, which asserts that the maximal amount of a flow (provided by the network) is equal to the capacity of a minimal cut, i.e.~a nontrivial partition of the graph node set $V$ into two parts such that the sum of the capacities of the edges connecting the two parts (the cut capacity) is minimum. Provided there is only a single sink, routing offers an optimal solution for transporting messages both when they are statistically independent~\cite{Lehman:04} and when they are generated by correlated sources~\cite{BarrosS:06}.

Max-flow min-cut arguments are useful also in the case of multicast networks, in which a single source
broadcasts a number of messages to a set of sinks. This network capacity problem was solved in~\cite{AhlswedeCLY:00}, where it is shown that applying coding operations at intermediate nodes (i.e.~{\it network coding}) is necessary to achieve the max-flow/min-cut bound of a general network. A converse proof for this problem, known as the {\it network information flow problem}, was provided by~\cite{Borade:02}, whereas linear network codes were proposed and discussed in~\cite{LiYC:03} and~\cite{KoetterM:03}.

\begin{figure}[p]
    \centering
        \includegraphics[width=4cm]{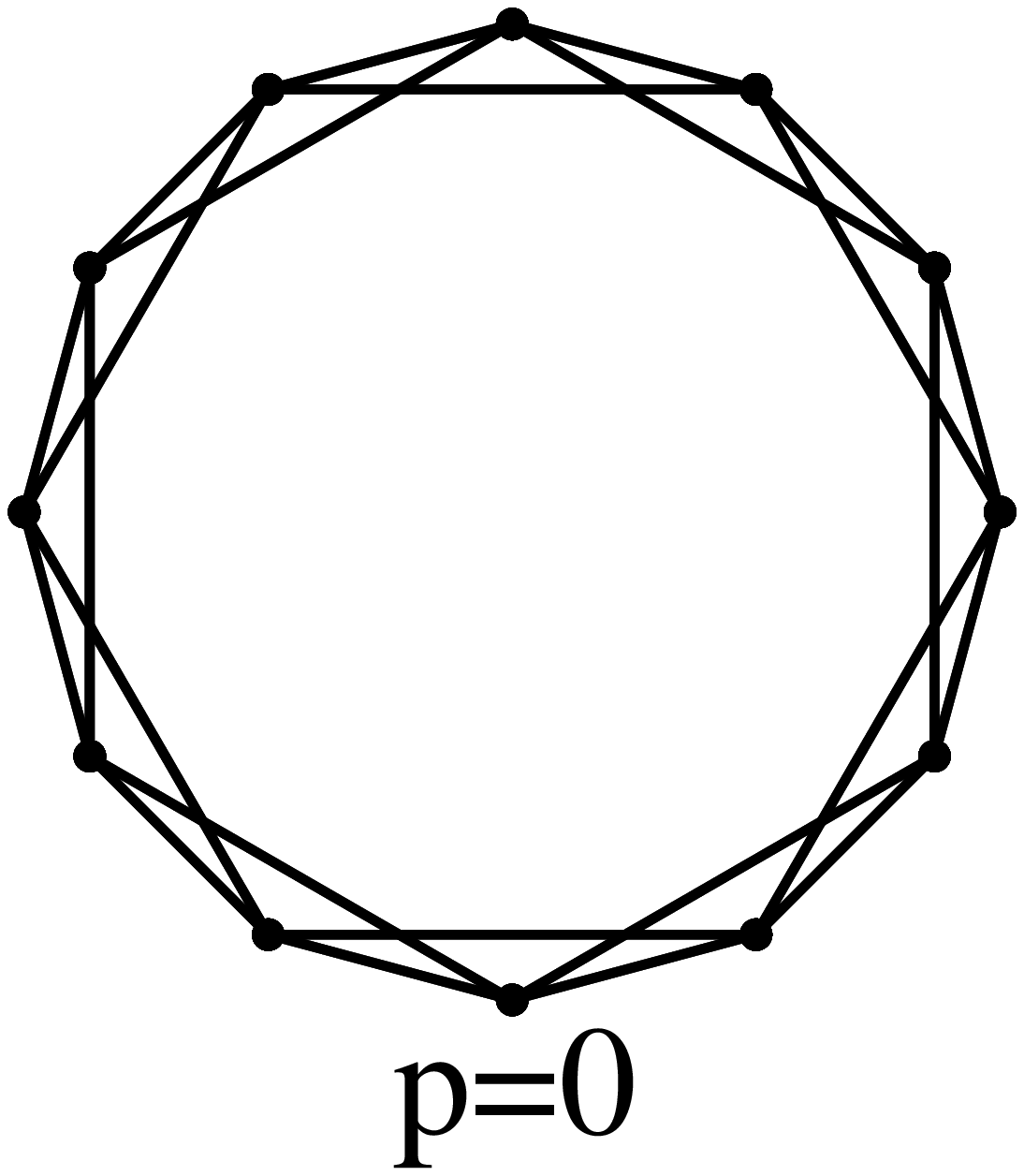}\hspace{0.35cm}
        \includegraphics[width=4cm]{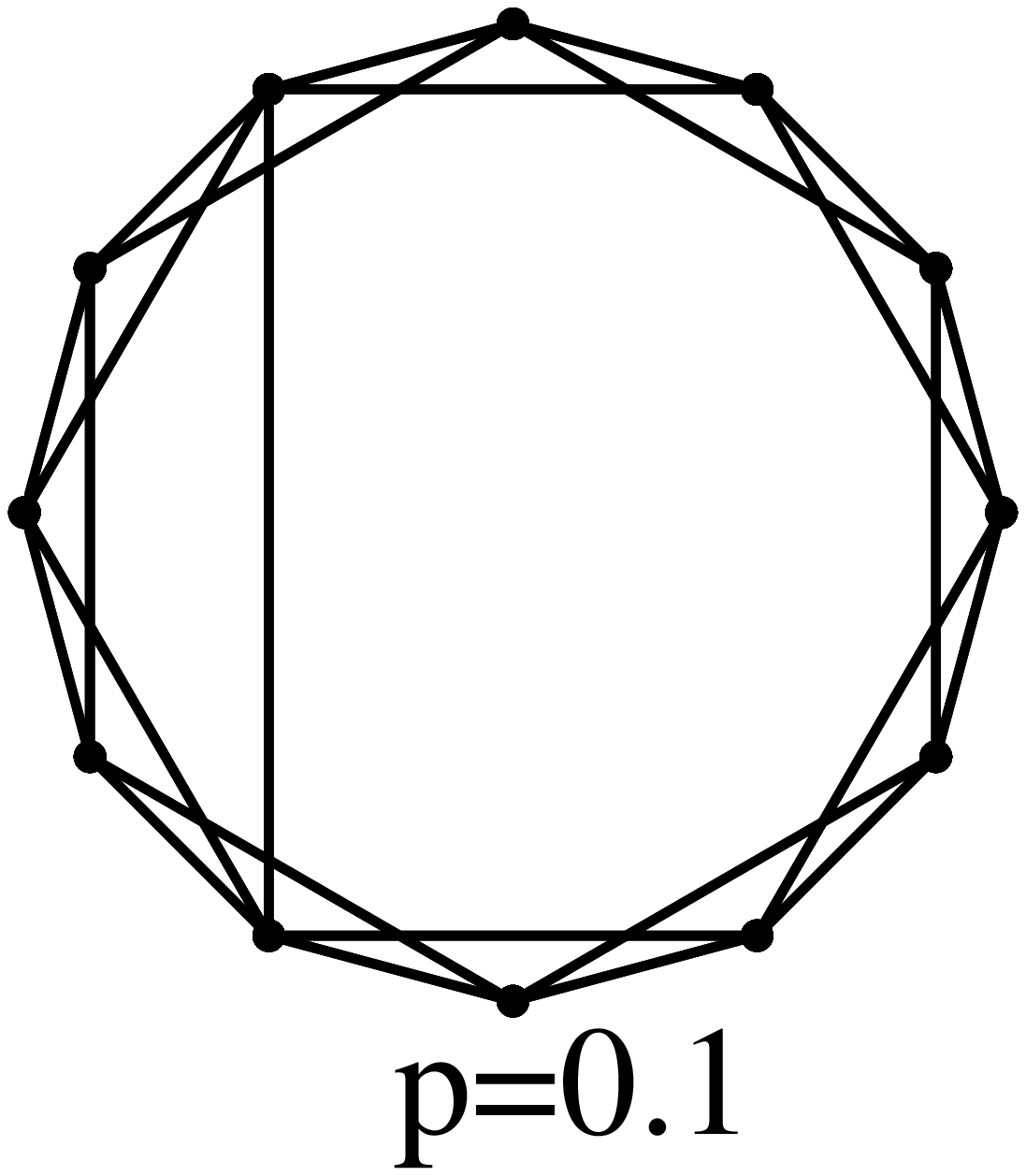} \hspace{0.25cm}
        \includegraphics[width=4cm]{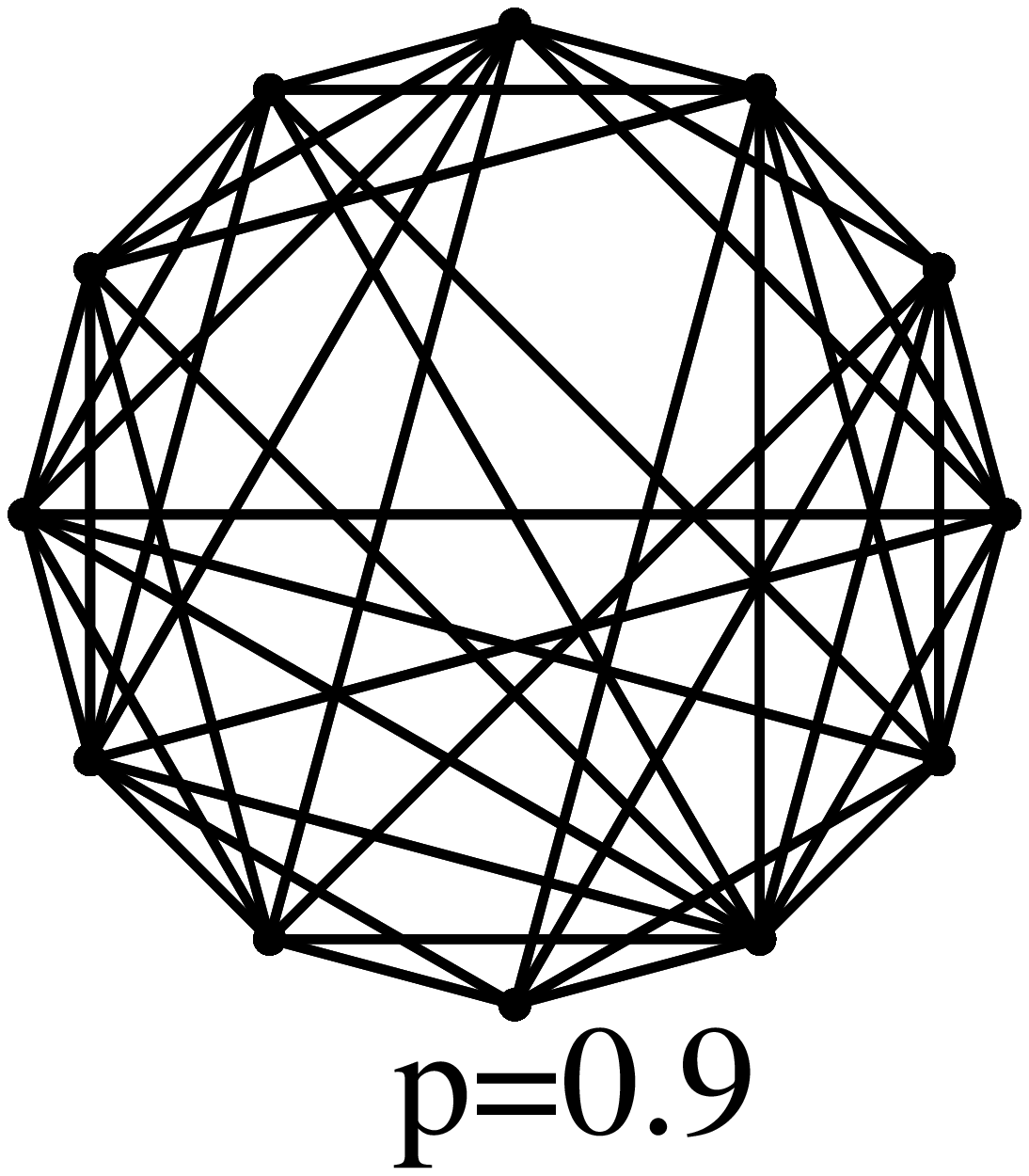}
    \caption{Small-world model with Shortcuts for different values of the adding probability $p$.}
    \label{fig:interpolation1}
\end{figure}

\begin{figure}[p]
    \centering
        \includegraphics[width=4cm]{k4p0.eps}\hspace{0.35cm}
        \includegraphics[width=4cm]{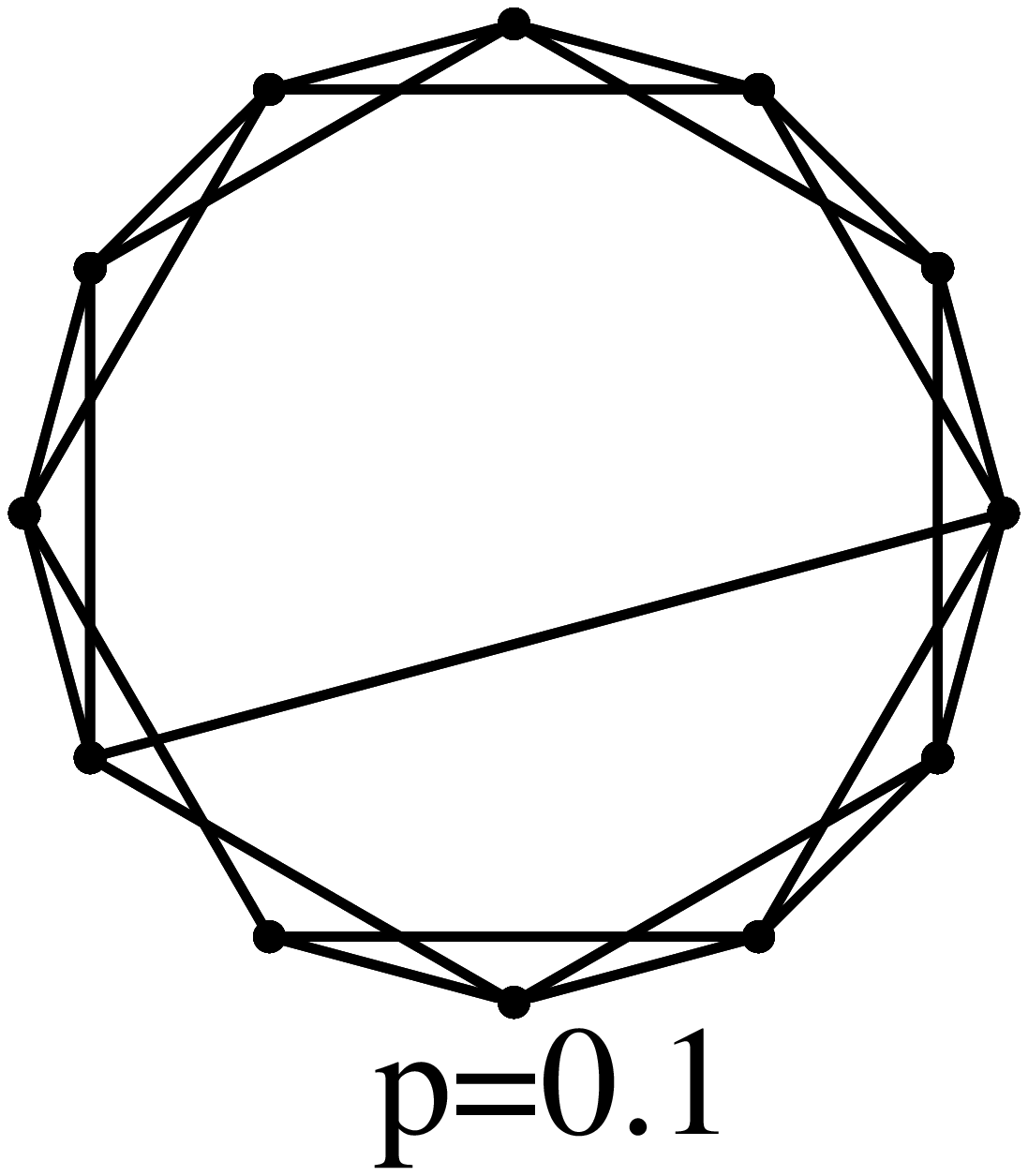} \hspace{0.25cm}
        \includegraphics[width=4cm]{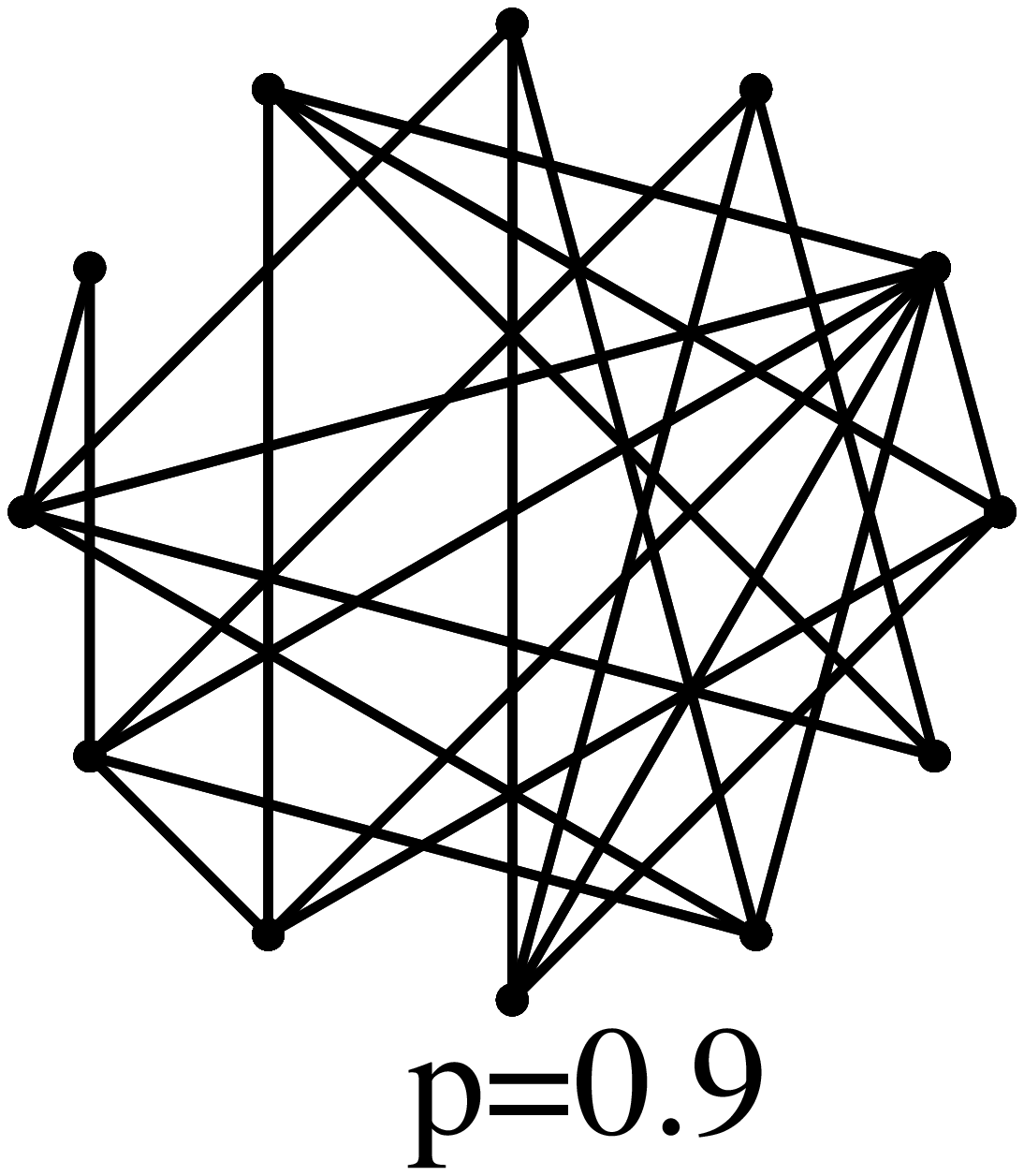}
    \caption{Small-world model with rewiring for different values of the rewiring probability $p$.}
    \label{fig:interpolation2}
\end{figure}

When the topology of the network is modeled by a randomly constructed graph, the natural goal is a probabilistic characterization of the minimum cut, which in the spirit of the network information flow literature~\cite{AhlswedeCLY:00}  we shall sometimes call (admittedly with some abuse) the {\it capacity} of the random network. Although some capacity results of this flavor are available for particular instances, most notably for  Erd\"os-R\'enyi graphs and random geometric graphs~\cite{ramamoorthy:05}, the problem remains open for many relevant classes of random graphs. Motivated by this observation, we make the following contributions:
\begin{itemize}
\item {\it A Max-flow Min-cut Theorem:} We introduce the {\it independence-in-cut} property, which is satisfied by large classes of random graphs, and derive inner and outer bounds for the minimum cut of any network that possesses this basic property.  In contrast with~\cite{ramamoorthy:05}, our approach is based on Hoeffding's inequality, which allows us to prove a theorem that is valid for a larger class of networks.
\item {\it Capacity Bounds for Small-World Networks:} Based on the aforementioned max-flow min-cut theorem, we are able to characterize the max-flow min-cut capacity of Small-World networks with shortcuts and with rewiring~\cite{watts:98}. Our results show, somewhat surprisingly, that, up to a constant factor, a rewiring rule that preserves the independence-in-cut property does not affect the capacity of large small-world networks.
\item {\it Capacity Bounds for Dual Radio Networks:} We are able to apply our theorem also to wireless network models in which some of the nodes are able to establish both short-range and long-range connections by means of dual radio interfaces. The capacity bounds thus obtained shed some light on the potential gains of this technology.
\end{itemize}
Our motivation to consider small-world networks, i.e. graphs with high clustering coefficients and small average path length, stems from their proven ability to capture fundamental properties of relevant phenomena and structures in sociology, biology, statistical physics and man-made networks. Beyond well-known examples such as Milgram's ''six degrees of separation"~\cite{milgram:67} between any two people in the United States and the Hollywood graph with links between actors, small-world structures appear in such diverse networks as the U.S. electric power grid, the nervous system of the nematode worm {\it Caenorhabditis elegans}~\cite{yamamoto:92}, food webs~\cite{martinez:00}, telephone call graphs~\cite{westbrook:98}, and, most strikingly,  the World Wide Web~\cite{broder:00}. The term small-world graph itself was coined by Watts and Strogatz, who in their seminal paper ~\cite{watts:98} defined a class of models which interpolate between regular lattices and random Erd\"os-R\'enyi graphs by adding shortcuts or rewiring edges with a certain probability $p$ (see Figures \ref{fig:interpolation1} and \ref{fig:interpolation2}). The most striking feature of these models is that for increasing values of $p$ the average shortest-path length diminishes sharply, whereas the clustering coefficient, defined as the expected value of the number of links between the neighbors of a node divided by the total number of links that could exist between them, remains practically constant during this transition.
Since small-world graphs were first proposed as models for complex
networks~\cite{watts:98} and ~\cite{NewmanW:99}, most contributions have focused essentially on connectivity parameters such as the degree distribution, the clustering coefficient or the shortest path length between two nodes (see e.g.~\cite{strogatz:01}). In spite of its arguable relevance --- particularly where communication networks are concerned --- the {\it capacity} of small-world networks has, to the best of our knowledge, not yet been studied in depth by the scientific community.

The second class of networks addressed in this paper is motivated by the fact that wireless interfaces become standard commodities and communication devices with multiple radio interfaces appear in various products.
Thus, it is only natural to ask whether the aforementioned devices can lead to
substantial performance gains in wireless communication networks.
Promising examples include~\cite{MultiRadio:04}, where
 multiple radios are used to provide  better
performance and greater functionality for users,
and ~\cite{PerRag:05}, where it is shown that using radio
hierarchies can reduce power consumption.
This growing interest in wireless systems with multiple radios
(for example, a Bluetooth interface and an IEEE
802.11 wi-fi card) motivates
us to study the impact of dual radio devices on the  capacity
of wireless networks.
%

The rest of the paper is organized as follows. \secref{sec:main} states the problem and proves our main theorem. The results for small-world networks and dual radio networks then follow in \secref{sec:swn} and \secref{sec:drn}, respectively. The paper concludes with \secref{sec:conclusions}.

\section{Main Result}
\label{sec:main}

Consider a graph $G=(V,E)$, where $V$ represents the set of nodes of the graph and $E$ the set of edges connecting these nodes. In the rest of the paper, we assume that the edges in the graph represent communication links with unitary capacity.

\begin{definition}
Consider a graph $G=(V,E)$ with $|V|=n$, a source $s$, a set $T$ of terminals and a set $R$ of relay nodes such that $V=\{s\} \cup R \cup T$. Let $t$ be a terminal node, i.e. $t\in T$ and let $N$ be the number of relay nodes, i.e. $N=|R|=n-1-|T|$. A $s$-$t$-cut of size $x$ in the graph $G$ is a partition of the set of relay nodes $R$ into two sets $V_k$ and $\overline{V}_k$ such that
$|V_k|=x$ and $|\overline{V}_k|=N-x$, $R=V_x\cup \overline{V}_x$ and $V_x\cap \overline{V}_x =\emptyset$.
The edges {\it crossing} the cut are given by
$E\cap \{ (s,i):i\in \overline{V}_x \}$,
$E\cap \{(j,t): j\in V_x \}$,
and $E\cap \{ (j,i): j\in V_x, i \in \overline{V}_x \}$.
\end{definition}

The following definition describes the capacity of a cut as the sum of the capacities of the edges crossing the cut.
\begin{definition}
Consider a graph $G=(V,E)$ and a $s$-$t$-cut of size $x$, with the corresponding sets $V_x$ and $\overline{V}_x$. The {\it capacity of the cut}, denoted by $C_x$, is given by $C_x=\sum\limits_{i\in \overline{V}_x}C_{si}+\sum\limits_{j\in V_x}\sum\limits_{i\in \overline{V}_x}C_{ji}+\sum\limits_{j\in V_x}C_{jt},$ where $C_{ij}$ denotes the capacity of the link between nodes $i$ and $j$.
\end{definition}

In the spirit of~\cite{ford:62}, we will refer to the value of the minimum $s$-$t$-cut as the $s$-$t$-{\it capacity}, denoted by $C_{s;t}$. In the case of multiple terminals, denoting by $T$ the set of terminals, the $s$-$T$-capacity, denoted by $C_{s;T}$, is the minimum of the $s$-$t$-capacities over all terminals, i.e. $C_{s;T}=\min_{t\in T} C_{s;t}$.

\begin{definition}
\label{defi:inde}
We say that a graph $G$ has the {\it independence-in-cut} property if, for every cut in the graph $C_x=\sum\limits_{i\in \overline{V}_x}C_{si}+\sum\limits_{j\in V_x}\sum\limits_{i\in \overline{V}_x}C_{ji}+\sum\limits_{j\in V_x}C_{jt}$, we have
$\prob (C_x=c_x)=\prod_{i\in \overline{V}_x} \prob(C_{si}=c_{si}) \cdot \prod_{j\in V_x}\prod_{i\in \overline{V}_x} \prob (C_{ji}=c_{ji})\cdot\prod_{j\in V_x}\prob(C_{jt}=c_{jt}),$ i.e. all the variables in the sum are independent random variables.
\end{definition}

Notice that, based on this definition, a graph with the independence-in-cut property is not necessarily a graph in which all the edges in the graph are independent random variables. An example of this is the case of Dual Radio Networks, discussed in detail in \secref{sec:drn}, where we shall show that, although there is dependency between some edges, we have that, given a cut, all the edges crossing it are independent random variables. This observation is valid for every cut.

In our approach, we make use of the independence-in-cut property to compute bounds on the probability that the capacity of a cut is close to its expected value. In fact, we shall view a cut as the sum of random variables. Given the independence-in-cut property, these variables are independent, which allows us to provide the desired bounds. If these variables are not independent, i.e. if the edges that cross a given cut are not independent random variables, the computation of these bounds becomes extremely difficult, since we are required to bound the sum of correlated random variables, irrespective of the correlation structure.

Capacity bounds for Erd\"os-R\'enyi graphs and random geometric graphs are the main focus of~\cite{ramamoorthy:05}. The bounds were derived specifically for each of the models, using Chernoff techniques. This limits the analysis to the case of networks with independent and identically distributed edges. We shall use some of the techniques presented in ~\cite{ramamoorthy:05}, but instead of Chernoff bounds our approach exploits Hoeffding's inequality~\cite{hoeffding:63} to derive a more general result. The resulting theorem is true for any network that verifies the independence-in-cut property --- edges are not required to be identically distributed.

Our main result is given by the following theorem.
\begin{theorem}
\label{th:main}
Consider a graph $G=(V,E)$ (with $n=|V|$) with the independence-in-cut property. Consider also a source $s$ and a set $T=\{t_1,\dots,t_{\alpha}\}$ of terminal nodes in $G$. Let $c_{\min}=\displaystyle \min_{C\in \cal{C}}E(C)$, where $\cal{C}$ is the set of all possible $s$-$t_1$-cuts. Let $\lambda= \displaystyle \min_{i,j:\prob(i\leftrightarrow j)>0}{\prob(i\leftrightarrow j)}$ and $\epsilon=\sqrt{\frac{d\ln (n-2)}{\lambda^2 (n-2) }}$ with $1<d<\frac{\lambda^2 (n-2)}{\ln (n-2)}$. The $s$-$T$-capacity, $C_{s;T}$, verifies
\begin{eqnarray*}
C_{s;T}&>&
(1-\epsilon)c_{\min} \textrm{ with probability }1-\textrm{O}\left(\frac{\alpha}{n^{2d}}
\right),\\
C_{s;T}&<&
(1+\epsilon)c_{\min} \textrm{ with probability } 1-\textrm{O}\left(\frac{1}{n^{2d}} \right).
\end{eqnarray*}
\end{theorem}

To be able to prove \theorref{th:main}, we first need to state and prove a few auxiliary results. We start by presenting a useful inequality.

\begin{lemma}[Hoeffding's inequality, from~\cite{hoeffding:63}]
\label{le:hoe}
For $X_1,X_2,\dots,X_m$ independent random variables with $\prob (X_i \in
[a_i,b_i])=1,\: \forall i\in \{1,2,\dots,m \}$, if we define
$S=X_1+X_2+\dots+X_m$, then $\prob (S-E(S)\geq mt)\leq \exp\left(-2m^2
t^2 / \sum\limits_{i=1}^{m}(b_i-a_i)^2\right).$
\end{lemma}

First, we determine an upper bound on the probability that the capacity of a cut takes a value much smaller than its expected value.

\begin{lemma}
\label{le:first}
Consider the single-source single-terminal case. For $\epsilon>0$ and
$N\geq 2$, we have that $\prob (C_x\leq (1-\epsilon)\expectation{C_x} )\leq
e^{-2(N+x(N-x))\epsilon^2 \lambda^2 }.$
\end{lemma}

\begin{proof}
We start by writing\vspace{-0.4cm}
\begin{eqnarray}
\label{hoe}
\prob  \left[C_x\leq (1-\epsilon)E(C_x)\right]
=  \prob \left[-C_x-E(-C_x)\geq \epsilon E(C_x)\right].
\end{eqnarray}

\vspace{-0.4cm}To compute the desired upper bound, we shall use the Hoeffding's inequality
(\lemmaref{le:hoe}). We have that $C_x=\sum\limits_{i\in \overline{V}_x}C_{si}+\sum\limits_{j\in V_x}\sum\limits_{i\in \overline{V}_x}C_{ji}+\sum\limits_{j\in V_x}C_{jt}$. We have that $C_{ab}\in \{0,1\}$ and $E(C_{ab})=\prob (a\leftrightarrow b)$. Therefore \vspace{-0.4cm}$$E(C_x)\stackrel{(a)}{\geq} (N-x+x(N-x)+x)\lambda=(N+x(N-x))\lambda,\vspace{-0.2cm}$$
where (a) follows from setting $\lambda= \displaystyle \min_{i,j:\prob(i\leftrightarrow j)>0}{\prob(i\leftrightarrow j)}$. Thus, we have that if $-C_x-E(-C_x)\geq \epsilon E(C_x)$, then $-C_x-E(-C_x)\geq (N+1+x(N-x))\epsilon \lambda$. Therefore, \vspace{-0.3cm}$$\prob \left[-C_x-E(-C_x)\geq \epsilon E(C_x)\right]\leq \prob \left[-C_x-E(-C_x)\geq (N+1+x(N-x))\epsilon \lambda\right].\vspace{-0.3cm}$$ Moreover, from \eqref{hoe},\vspace{-0.4cm}
\begin{eqnarray}
\label{aux}
\prob  \left[C_x\leq (1-\epsilon)E(C_x)\right] \leq \prob \left[-C_x-E(-C_x)\geq (N+1+x(N-x))\epsilon \lambda\right] .
\end{eqnarray}

\vspace{-0.4cm}Now, because the graph $G$ has the independence-in-cut property, $C_x$ can be viewed as the sum of $N+x(N-x)$ independent Bernoulli distributed random variables. Therefore, we can apply \lemmaref{le:hoe} to \eqref{aux}, with $m=N+x(N-x)$ and $t=\epsilon \lambda$, and we get
\begin{eqnarray}
\label{conc}
\prob  (C_x\leq (1-\epsilon)E(C_x)) &\leq& \exp \left( \frac{-2(N+x(N-x))^2\epsilon^2 \lambda^2}{N+x(N-x)} \right) \nonumber= \exp \left( -2(N+x(N-x))\epsilon^2 \lambda^2 \right).
\end{eqnarray}
\end{proof}

\begin{remark}
The previous result, \lemmaref{le:first}, is valid for networks where the edges represent links with unitary capacity. In \lemmaref{le:hoe} this corresponds to the case in which all variables $X_i$ lie in the unit interval $[0,1]$. This result (and consequently \theorref{th:main}) can be easily extended for networks with different edge capacities, because Hoeffding's inequality is valid for variables $X_i\in [c_i,b_i]$, in general. For simplicity, we consider only the case of edges with unitary capacity.
\end{remark}

Using the previous result, we obtain another useful inequality.

\begin{cor}
\label{cor:ak}
Let $A_x$ be the
event given by $\{ C_x < (1-\epsilon)\expectation{C_x}\}$. Then, $\prob (\cup_x A_x)\leq
2e^{-2\epsilon^2 \lambda^2 N} \cdot \left[1+e^{-\epsilon^2 \lambda^2 N}\right]^N.$
\end{cor}

\begin{proof}
By \lemmaref{le:first}, we have that $\prob(A_x)\leq
e^{-2(N+x(N-x))\epsilon^2 \lambda^2}$. Notice that, for each $x\in \{0,...,N\}$,
there are ${N \choose
x}$ cuts in which one of the partitions consists of $x$ nodes and the source.
Therefore, we can write
\begin{eqnarray}
\label{eq:nsei22}
\prob (\cup_x A_x) &\leq& \sum \limits_{x=0}^{N} {N \choose x}\prob (A_x)\ \leq \sum \limits_{x=0}^{N} {N \choose x} e^{-2(N+x(N-x))\epsilon^2 \lambda^2}.
\end{eqnarray}

We have that\vspace{-0.4cm}
\begin{eqnarray*}
e^{-2(N+x(N-x))\epsilon^2 \lambda^2} &=& e^{-2\epsilon^2 \lambda^2 N -2\epsilon^2 \lambda^2 x(N-x)}= e^{-2\epsilon^2 \lambda^2 N}\cdot e^{-2\epsilon^2 \lambda^2 N \cdot N \frac{x}{N}\left(1-\frac{x}{N} \right)}.
\end{eqnarray*}
Setting $\beta=e^{-2\epsilon^2 \lambda^2 N}$, we get $e^{-2(N+x(N-x))\epsilon^2 \lambda^2}=\beta \cdot \beta^{N\frac{x}{N}(1-\frac{x}{N})}$. From \eqref{eq:nsei22}, we get
\begin{eqnarray*}
\prob (\cup_x A_x)  &\leq&  \beta \sum
\limits_{x=0}^{N} {N \choose x} \beta^{N\frac{x}{N}(1-\frac{x}{N})}
= \beta   \left(  \sum
\limits_{x=0}^{\lfloor N/2 \rfloor}  {N \choose x}
\beta^{N\frac{x}{N}(1-\frac{x}{N})}  +  \sum
\limits_{x=\lfloor N/2 \rfloor+1}^{N}  {N \choose x}
\beta^{N\frac{x}{N}(1-\frac{x}{N})} \right) .
\end{eqnarray*}
Notice that, when $\frac{x}{N}\in [0,1/2]$, we have $\:\frac{x}{N}(1-\frac{x}{N})\geq
\frac{x}{2N},$ and when $\frac{x}{N}\in [1/2,1],$ we get $\: \frac{x}{N}
(1-\frac{x}{N})\geq \frac{N-x}{2N}.$ Thus, we can write
\begin{eqnarray*}
\prob (\cup_x A_x)  &\leq& \beta
\left(  \sum \limits_{x=0}^{\lfloor N/2 \rfloor}
{N \choose x} \beta^{N\frac{x}{2N}} +  \sum \limits_{x=\lfloor N/2
\rfloor+1}^{N}  {N \choose x}
\beta^{N\frac{N-x}{2N})}\right)\\
&=&  \beta\left( \sum \limits_{x=0}^{N} {N \choose x}
\left(\beta^{\frac{1}{2}}\right)^x + \sum \limits_{x=0}^{N} {N \choose x} \left(
\beta^{\frac{1}{2}}\right)^{N-x}\right)\\
&\stackrel{(a)}{=}&  2\beta (1+\sqrt{\beta})^N
\stackrel{(b)}{=}   2e^{-2\epsilon^2 \lambda^2 N}\cdot
\left[1+e^{-\epsilon^2 \lambda^2 N}\right]^N,
\end{eqnarray*}
where we use the following arguments:
\begin{itemize}
\item[(a)] follows from the fact that $(y+z)^m=\sum\limits_{x=0}^{m} {m \choose
x} y^x z^{m-x}$, thus taking $m=N$, $y=\beta^{\frac{1}{2}}=\sqrt{\beta}$ and $z=1$, we get $\sum\limits_{x=0}^{N} {N \choose x} \left(\beta^{\frac{1}{2}} \right)^x=(1+\sqrt{\beta})^N$ and taking $m=N$, $y=$1 and $z=\beta^{\frac{1}{2}}=\sqrt{\beta}$, we get $\sum\limits_{x=0}^{N} {N \choose x} \left(\beta^{\frac{1}{2}} \right)^{N-x}=(1+\sqrt{\beta})^N;$
\item[(b)] follows from substituting $\beta$ by
$e^{-2\epsilon^2 \lambda^2 N}$.
\end{itemize}
\vspace{-0.5cm}
\end{proof}

Now, using \corolref{cor:ak}, we can bound the global minimum cut of $G_s$.

\begin{cor}
\label{cor:cmin}
For all $t_i\in T$, we have that $\prob (C_{s,t_i}  \leq
 (1 - \epsilon)c_{\min})  \leq
 2e^{-2\epsilon^2 \lambda^2 N}\cdot
\left[1+e^{-\epsilon^2 \lambda^2 N}\right]^N .$
\end{cor}

\begin{proof}
Let $\tilde{A}_x$ be the event $\{ C_x<(1-\epsilon)c_{\min} \}$ and let
$A_x$ be the event $\{ C_x<(1-\epsilon)\expectation{C_x} \}$. We have that
$E(C_x)\geq c_{\min},\: \forall x \in {0,...,N} $, because $c_{\min}=\displaystyle \min_{C\in \cal{C}}E(C)$, where $\cal{C}$ is the set of all possible $s$-$t_i$-cuts. Consequently, we have $\tilde{A}_x \subseteq A_x$,
which implies that $\cup_x \tilde{A}_x \subseteq \cup_x A_x$, resulting in
$
\prob (C_{s,t_i}\leq (1-\epsilon)
c_{\min})= \prob (\cup_x \tilde{A}_x)
\leq \prob (\cup_x A_x).
$
Applying \corolref{cor:ak} concludes the proof.
\end{proof}

We are now ready to prove \theorref{th:main}.

\begin{proofmain}
Replacing $\epsilon$ in \corolref{cor:cmin} by the
expression $\sqrt{\frac{d\ln (n-2)}{\lambda^2 (n-2)}}=\sqrt{\frac{Nd\ln
N}{\lambda^2 N}}$, we obtain\vspace{-0.4cm}
\begin{eqnarray*}
\prob (C_{s,t_i}  \leq
 (1 - \epsilon)c_{\min})
 &\leq&  2e^{-2\lambda^2 N\frac{d\ln
N}{\lambda^2 N}} \cdot [1 +
e^{-\lambda^2 N\frac{d\ln N}{\lambda^2 N}}]^N = 2e^{-2d \ln
N} \cdot [1 +
e^{-d \ln N}]^N\\
&=& 2e^{\ln(
N^{-2d})} \cdot [1 +
e^{\ln (N^{-d })}]^N
=  \frac{2}{N^{2d}}\cdot \left[1+\frac{1}{N^{d}}\right]^N.
\end{eqnarray*}
Using once again the fact that $(y+z)^N=\sum\limits_{x=0}^{N} {N \choose x} y^x z^{N-x}$, we get
$\left[1+\frac{1}{N^{d}}\right]^N=\sum\limits_{x=0}^{N} {N \choose x}
\left(\frac{1}{N^d} \right)^x.$ Therefore, we have that\vspace{-0.4cm}
\begin{eqnarray}
\prob (C_{s,t_i}  \leq
 (1 - \epsilon)c_{\min})  &\leq&
  \frac{2}{N^{2d}}\cdot \sum\limits_{x=0}^{N} {N \choose x}
\left(\frac{1}{N^d} \right)^x\nonumber \\
&\stackrel{(a)}{\leq}&  \frac{2}{N^{2d}}\cdot
\sum\limits_{x=0}^{\infty} \left(\frac{N}{N^d} \right)^x
\stackrel{(b)}{=}  \frac{2}{N^{2d}-N^{d+1}}
\approx  O\left(\frac{1}{N^{2d}} \right)\nonumber
=  O\left(\frac{1}{n^{2d}} \right) \label{eq:singlesource}
\end{eqnarray}
where we used the following arguments:
\begin{itemize}
\item[(a)] follows from the fact that ${N \choose
x}=\frac{N!}{(N-x)!x!}=\frac{N\cdot(N-1)\cdot \ \dots \ \cdot (N-x+1)}{x!}$, thus
${N \choose x}\leq N\cdot(N-1)\cdot \ \dots \ \cdot (N-x+1)\leq N^x$;
\item[(b)] follows from the fact that $\sum\limits_{x=0}^{\infty}y^x
=\frac{1}{1-y}$, for $|y|<1$, leading to (for $d>1$) $\sum\limits_{x=0}^{\infty}
\left(\frac{N}{N^d} \right)^x = \frac{1}{1-N^{1-d}},$ which implies that
$
\frac{2}{N^{2d}}\cdot \sum\limits_{x=0}^{\infty} \left(\frac{N}{N^d}
\right)^x=\frac{2}{N^{2d}-N^{d+1}}.
$
\end{itemize}

Using the bounds we have already constructed for the single-source single-terminal case, we can easily obtain bounds for the single-source multiple-terminals case. Since $C_{s;T}=\displaystyle \min_{t_i\in T} C_{s;t_i}$, we have that\vspace{-0.4cm}
\begin{eqnarray}
\prob(C_{s;T} \leq (1-\epsilon)c_{\min}) &=& \prob \left( \bigcup_{t_i\in T} \{ C_{s;t_i} \leq (1-\epsilon)c_{\min} \}\right)\nonumber
\leq \sum_{t_i\in T} \prob(C_{s;t_i} \leq (1-\epsilon)c_{\min}).\label{eq:cormain}
\end{eqnarray}

By \theorref{eq:singlesource}, we have that $\prob(C_{s;t_i} \leq (1-\epsilon)c_{\min})=\textrm{O}\left(\frac{1}{n^{2d}}
\right),\: \forall t_i \in T$. Thus, from \eqref{eq:cormain}, we have that $\prob(C_{s;T} \leq (1-\epsilon)c_{\min})=\textrm{O}\left(\frac{\alpha}{n^{2d}}
\right).$

Now, to compute the upper bound on $\prob (C_{s;T} \geq
(1+\epsilon)c_{\min})$, let $C^*$ be a cut such that $E(C^*)=c_{\min}$. Notice that, by definition, any cut is greater or equal to $C_{s;T}$, in particular the cut $C^*$. Thus, if $C_{s;T}\geq(1+\epsilon)c_{\min}$ then $C^* \geq (1+\epsilon)c_{\min}$. Therefore, $\prob ( C_{s;T}\geq
(1+\epsilon)c_{\min} ) \leq \prob (C^* \geq (1+\epsilon)c_{\min}),$ which is equivalent to \vspace{-0.4cm}
\begin{eqnarray}
\prob ( C_{s;T}\geq (1+\epsilon)c_{\min}) \leq \prob (C^*-c_{\min} \geq \epsilon c_{\min}).\label{nsei5}
\end{eqnarray}

\vspace{-0.4cm}Define $\delta$ as the number of random variables that define the cut $C^*$, i.e. $\delta$ is the number of edges that possibly cross the cut $C^*$. Because $\lambda= \displaystyle \min_{i,j:\prob(i\leftrightarrow j)>0}{\prob(i\leftrightarrow j)}$, we have that $c_{\min}\geq \delta \lambda$. Thus, if $C^*-c_{\min} \geq \epsilon c_{\min}$, then $C^*-c_{\min} \geq \epsilon \delta \lambda$. Hence
$\prob (C^*-c_{\min} \geq \epsilon c_{\min}) \leq \prob ( C^*-c_{\min} \geq \epsilon \delta \lambda )$
and, from \eqref{nsei5},\vspace{-0.4cm}
\begin{eqnarray}
\prob (C_{s;T}\geq (1+\epsilon)c_{\min}) \leq \prob ( C^*-c_{\min} \geq \epsilon \delta \lambda )\label{nsei4}
\end{eqnarray}

\vspace{-0.4cm}Because the graph $G$ has the independence-in-cut property, $C^*$ can be viewed as the sum of $\delta$ independent and bounded random variables (more precisely, all the variables belong to the interval $[0,1]$). Thus, noticing that $E(C^*)=c_{\min}$ and applying Hoeffding's inequality (\lemmaref{le:hoe}) with $m=\delta$, $S=C^*$ and $t=\epsilon \lambda$, we have that
$$\prob ( C^* -c_{\min}
\geq \epsilon c_{\min} )\leq \exp \left(
-\frac{2\delta^2 \epsilon^2 \lambda^2}{\delta} \right)=\exp \left(
-2\delta \epsilon^2 \lambda^2 \right).$$

Recall that in a $s$-$t_i$-cut of size $x$ there are $N+x(N-x)$ random variables (with $N=n-2$). Since $x\in \{0,\dots,N \}$, we have that the number of random variables that define a cut is at least $N$, and this is true for every cut. Thus, the same holds for the cut $C^*$, which implies that $\delta \geq N$. This is equivalent to $\delta \geq n-2$, hence \vspace{-0.4cm}$$\prob ( C^* -c_{\min}
\geq \epsilon c_{\min} )\leq \exp \left(
-2(n-2) \epsilon^2 \lambda^2 \right),\vspace{-0.2cm}$$ and thus, from \eqref{nsei4}, we get $\prob (C_{s;T}\geq (1+\epsilon)c_{\min} )\leq \exp \left(
-2(n-2) \epsilon^2 \lambda^2 \right).$
Replacing $\epsilon$ by $\sqrt{\frac{d\ln (n-2)}{\lambda^2 (n-2) }}$, we obtain\vspace{-0.2cm}
\begin{eqnarray*}
\prob (C_{s;T}\geq (1+\epsilon)c_{\min} )&\leq& \exp \left(
-2(n-2) \frac{d\ln (n-2)}{\lambda^2 (n-2) } \lambda^2 \right)
= \exp (-2d \ln (n-2))
= \frac{1}{(n-2)^{2d}}\\
&=& \textrm{O}\left(\frac{1}{n^{2d}} \right).
\end{eqnarray*}
\end{proofmain}
\vspace{-0.8cm}
\proofend

\section{Small-World Networks}
\label{sec:swn}

\subsection{Classes of Small-World Networks}

\begin{figure}[p]
     \centering
        \includegraphics[width=5cm]{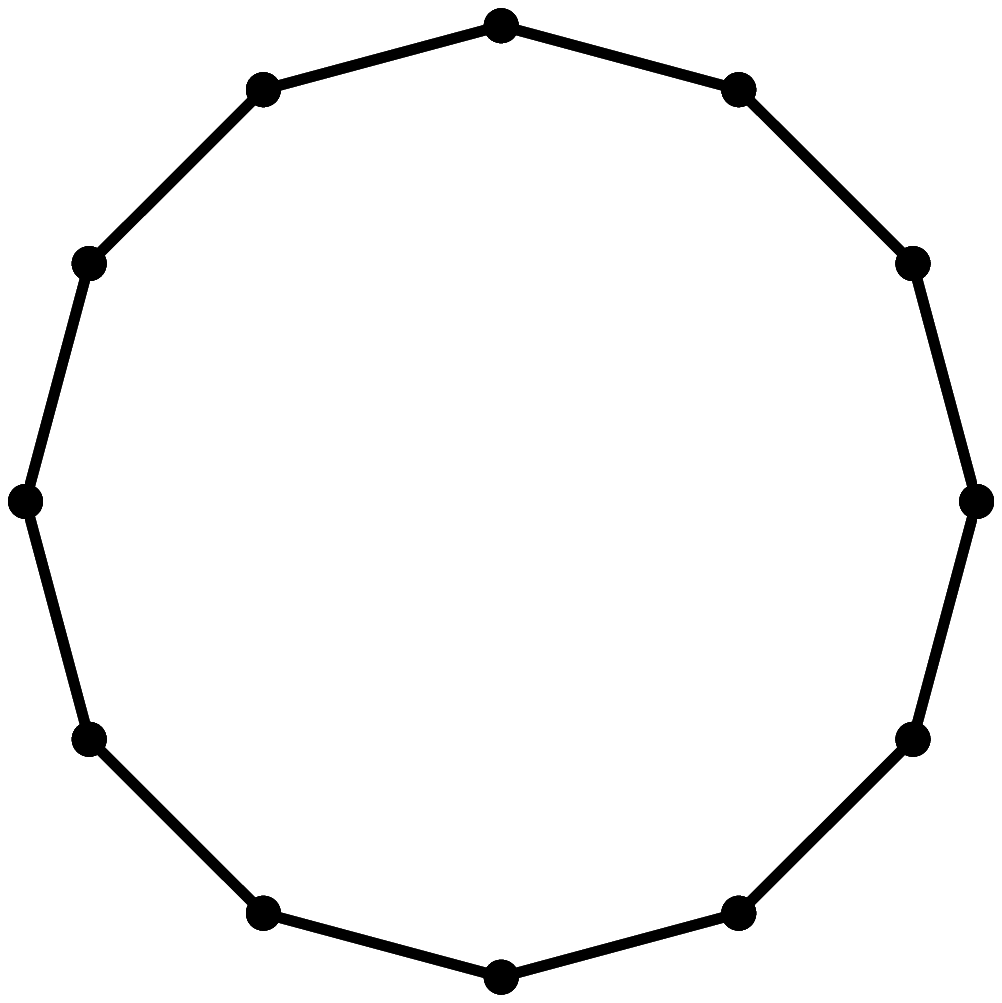}\hspace{-0.45cm}
        \includegraphics[width=5cm]{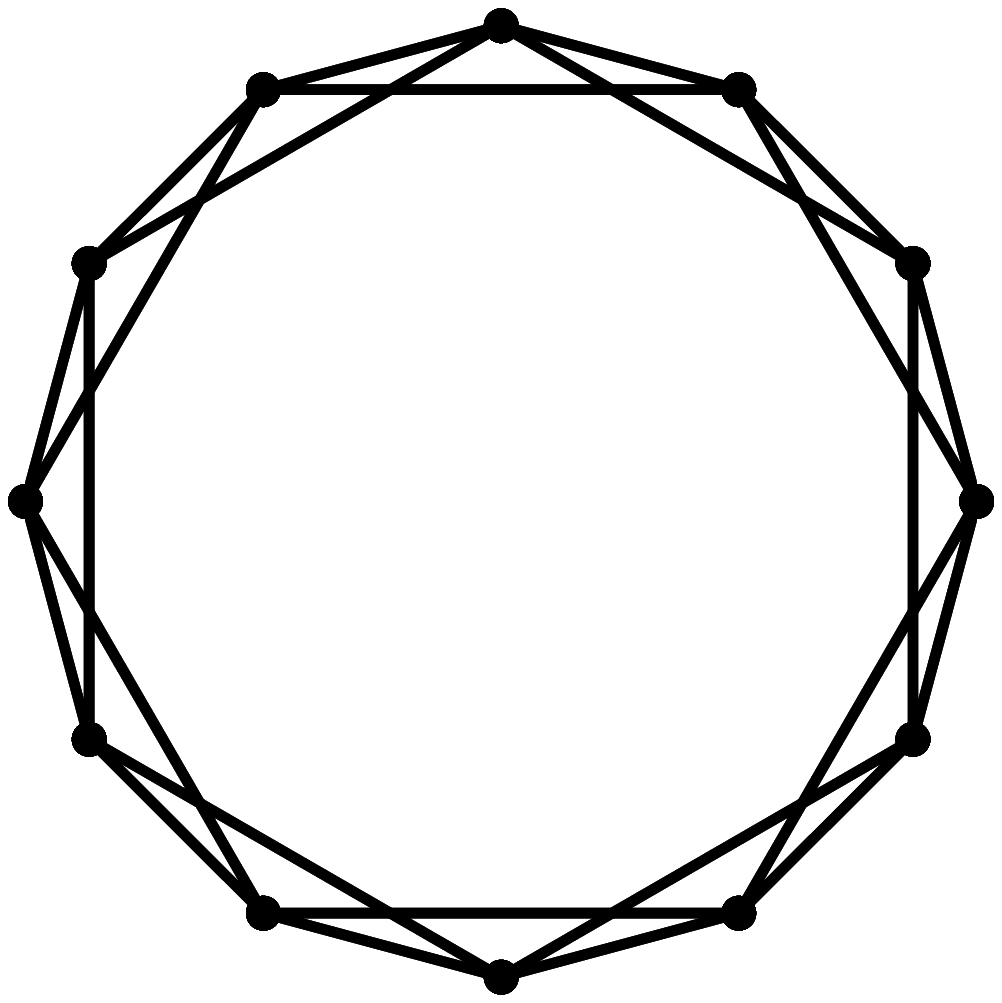} \hspace{-0.45cm}
        \includegraphics[width=5cm]{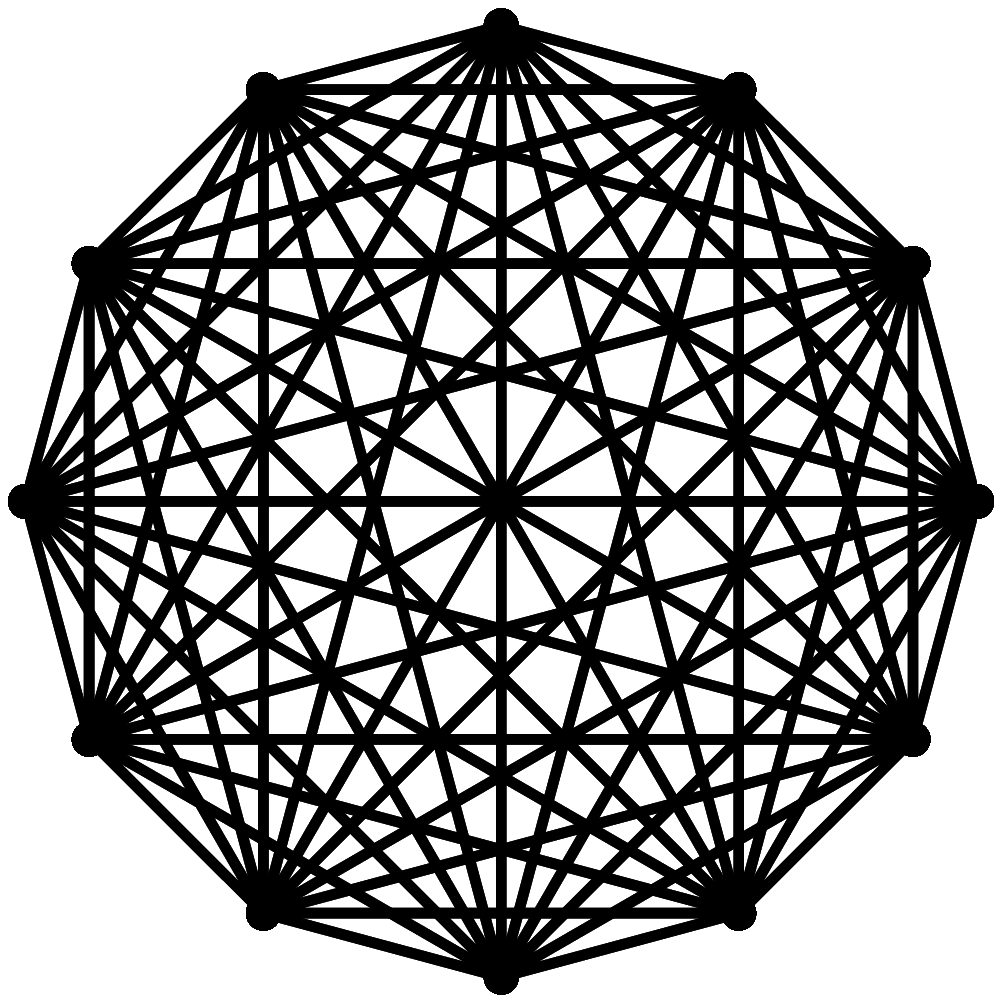}
    \caption{Illustration of a k-connected ring lattice: from left to right $k=2, 4, 12$.}
    \label{fig:ringlattice}
\end{figure}

We start by giving rigorous definitions for the classes of small-world networks under consideration. All the models in this paper are consider to be unweighted graphs containing no self-loops or multiple edges.
First, we require a precise notion of distance in a ring.
\begin{definition}
Consider a set of $n$ nodes connected by edges that form a ring
(see \fig{fig:ringlattice}, left plot). The {\it ring distance}
between two nodes is defined as the minimum number of {\it hops}
from one node to the other. If we number the nodes in clockwise
direction, starting from any node, then the ring distance between nodes $i$ and $j$ is given by $d(i,j)=min\{|i-j|,n-|i-j|\}.$
\end{definition}

For simplicity, we refer to $d(i,j)$ as the {\it distance} between $i$ and $j$. Next, we define a $k$-connected ring lattice, which serves as basis for the small-world models used in this paper.

\begin{definition}
A $k$-{\it connected ring lattice} (see \fig{fig:ringlattice}) is a graph $L=(V_L,E_L)$ with nodes $V_L$ and edges $E_L$, in which all nodes in $V_L$ are placed on a ring and are connected
to all the nodes within distance $\frac{k}{2}$.
\end{definition}
Notice that in a $k$-connected ring lattice, all the nodes have degree $k$. We are now ready to define the small-world models under consideration.

\begin{definition}[Small-World Network with Shortcuts~\cite{NewmanW:99}]
\label{def:shortcuts}
 We start with a $k$-connected ring lattice $L=(V_L,E_L)$ and let $E_C$ be the set of all possible edges between nodes in $V_L$. To obtain a {\it small-world network with shortcuts}, we add
to the ring lattice $L$ each edge $e\in E_C \backslash E_L$
with probability $p$.
\end{definition}

\begin{definition}[Small-World Network with Rewiring]
\label{def:watstro}
Consider a $k$-connected ring lattice $L=(V_L,E_L)$. To obtain a {\it small-world network with rewiring}, we proceed as follows. Let $E_R=E_L$ be the initial set of edges. Each edge $e \in E_L$ is removed from the set $E_R$ with probability $1-p$, where $p$ is called the {\it probability of
rewiring}. Each edge $e\notin E_L$ is then added to the set $E_R$ with probability $\frac{pk}{n-k-1}$. After considering all possible edges connecting nodes in $V_L$, the resulting small-world network is specified by
the graph $(V_L, E_R)$.
\end{definition}

This model is a variant of the small-world network with rewiring in~\cite{watts:98} in which all the edges can be viewed as independent random variables, thus satisfying the independence-in-cut property. Finding max-flow min-cut bounds for the original construction is intractable due to the complex dependencies between randomly rewired edges. To ensure the key property of constant average number of edges per node, as in~\cite{watts:98}, our definition attributes weight $\frac{pk}{n-k-1}$ to the edges that are not in the initial lattice. The expected number of edges per node in an instance of the model is thus given by $(1-p)k+\frac{pk}{n-k-1}(n-k-1)=k$.
\vspace{-0.4cm}

\subsection{Capacity Bounds for Small-World Networks}
We shall now use Theorem 1 to prove capacity bounds for the aforementioned small-world models. We start with a useful lemma.
\begin{lemma}
\label{lemma:latt}
Let $L=(V_L, E_L)$ be a $k$-connected ring lattice  and  let $G=(V_L,E)$ be a
fully connected graph (without self-loops), in which edges $e\in E_L$ have weight
$w_1\geq 0$ and edges $f\notin E_L$ have weight $w_2\geq 0$. Then, the global
minimum cut in $G$ is $k w_1+(n-1-k)w_2$.
\label{lemma:mincut}
\end{lemma}

\begin{proof}
We start by splitting $G$ into two subgraphs: a $k$-connected ring lattice $L$
with weights $w_1$ and a graph $F$ with nodes $V_L$ and all remaining edges of
weight $w_2$. Clearly, the value of a cut in $G$ is the sum of the values of the
same cut in $L$ and in $F$. Moreover, both in $L$ and in $F$, the global minimum
cut is a cut in which one of the partitions consists of one node (any other
partition increases the number of outgoing edges). Since each node in $L$ has
$k$ edges of weight $w_1$ and each node in $F$ has the remaining $n-1-k$ edges
of weight $w_2$, the result follows.
\end{proof}

The following theorem gives upper and lower bounds on the capacity of Small-World
Networks with Shortcuts.
\begin{theorem}
\label{th:adding}
The $s$-$T$-capacity of a Small-World Network with Shortcuts with parameters $n$, $p$ and $k$, denoted by $C_{s;T}^{\textrm{SWS}}$, satisfies the following inequalities:
\begin{eqnarray*}
C_{s;T}^{\textrm{SWS}}&>&
(1-\epsilon)[k+(n-1-k)p] \textrm{ with probability }1-\textrm{O}\left(\frac{\alpha}{n^{2d}}
\right)\\
C_{s;T}^{\textrm{SWS}}&<&(1+\epsilon)[k+(n-1-k)p] \textrm{ with probability } 1-\textrm{O}\left(\frac{1}{n^{2d}} \right),
\end{eqnarray*}
for $\epsilon=\sqrt{\frac{d\ln (n-2)}{p^2(n-2)}}$ and $1<d<\frac{p^2(n-2)}{\ln(n-2)}$. Moreover, $\displaystyle \lim_{n\rightarrow \infty} \epsilon =0$.
\end{theorem}

\begin{proof}
Consider a Small-World Network with Shortcuts $G_{\textrm{SWS}}=(V,E)$. Let $G_w$ be a fully-connected weighted simple graph with set of nodes $V$. If we assign to each edge $e=(i,j)$ in $G_w$ the weight $w_e=\prob \{ \textrm{$i$ and $j$ are connected in the Small-World with Shortcuts}\}$, we have that the expected value of a cut in $G_{\textrm{SWS}}$ is the value of the same cut in $G_w$. Therefore, since $c_{\min}=\displaystyle \min_{C\in \cal{C}}E(C)$, we have that $c_{\min}$ is the value of the minimum cut in the graph $G_w$. Notice that the weights are assigned as follows:
\begin{itemize}
\item The weight of the edges in the initial lattice of a Small-World Network
with Shortcuts is one (because they are not removed);
\item The weight of the remaining edges is $p$, (i.e.~the probability that an edge is added).
\end{itemize}
Therefore, $G_w$ is a graph in the conditions of \lemmaref{lemma:mincut}, with
$w_1=1$ and $w_2=p$. Hence, the global minimum cut in $G_w$ is given by $k+(n-1-k)p$, which is equivalent to $C_{\min}=k+(n-1-k)p$. Moreover, the minimum edge weight is $p$, i.e. $\lambda=p$. Therefore, using \theorref{th:main}, the bounds for the $s$-$T$-capacity of a Small-World Network with Shortcuts follow.
To conclude the proof, it just remains to notice that $\displaystyle \lim_{n\rightarrow \infty} \epsilon =\displaystyle \lim_{n\rightarrow \infty} \sqrt{\frac{d\ln (n-2)}{p^2(n-2)}} =0 .$
\end{proof}
We are also able to obtain a similar result for the case of rewiring, as previously defined.
\begin{theorem}
\label{th:rewi}
The $s$-$T$-capacity of a Small-World Network with Rewiring with parameters $n$, $p$ and $k$, denoted by $C_{s;T}^{\textrm{SWR}}$, satisfies the following inequalities:
\begin{eqnarray*}
C_{s;T}^{\textrm{SWR}}&>&
(1-\epsilon)k \textrm{ with probability }1-\textrm{O}\left(\frac{\alpha}{n^{2d}}
\right)\\
C_{s;T}^{\textrm{SWR}}&<&(1+\epsilon)k \textrm{ with probability } 1-\textrm{O}\left(\frac{1}{n^{2d}} \right),
\end{eqnarray*}
for $\lambda=\min \left\{ 1-p, \dfrac{pk}{n-k-1}\right\}$, $\epsilon=\sqrt{\frac{d\ln (n-2)}{\lambda^2(n-2)}}$ and $1<d<\frac{\lambda^2(n-2)}{\ln(n-2)}$. Moreover, if $p\geq 1-\frac{k}{n-1}$, then $\displaystyle \lim_{n\rightarrow \infty} \epsilon=0$. In the case of $p\leq 1-\frac{k}{n-1}$, if $\dfrac{k}{n}\geq \dfrac{1}{\ln^a(n)},\: \forall n\geq n_0$ for some $a>0$ and $n_0\in \mathbb{N}$, then $\displaystyle \lim_{n\rightarrow \infty} \epsilon=0$ and, if $\dfrac{k}{n}\leq \dfrac{b}{n},\: \forall n\geq n_1$ for some $b>0$ and $n_1\in \mathbb{N}$, then $\displaystyle \lim_{n\rightarrow \infty} \epsilon=\infty$.
\end{theorem}
\begin{proof}
As in the proof of \theorref{th:adding}, we consider a fully-connected weighted graph $G_w$ associated with a Small-World Network with Rewiring. From the definition of the model, we have that the weight of the edges $e\in E_L$ (i.e. the edges in the initial $k$-connected ring lattice) is given by $1-p$ and the weight of the remaining edges is given by $\frac{pk}{n-k-1}$. Notice that $G_w$ is a graph in the conditions of \lemmaref{lemma:mincut}, with
$w_1=1-p$ and $w_2=\frac{pk}{n-k-1}$. Therefore, the global minimum cut in $G_w$, is given by  $k(1-p)+(n-1-k)\frac{pk}{n-k-1}=k,$ which, using similar arguments to those in the proof of \theorref{th:adding}, is equivalent to $c_{\min}=k$.

We have that there are only two different probability values: $1-p$ and $\frac{pk}{n-k-1}$. Therefore, $\lambda=\min \left\{ 1-p, \dfrac{pk}{n-k-1}\right\}$. Notice also that all the edges are independent random variables (by the definition of the model). Hence, using \theorref{th:main}, we can obtain the sought bounds.
We can write
$1-\frac{k}{n-1} \geq p
\Leftrightarrow \frac{n-k-1}{n-1} \geq p
\Leftrightarrow n-k-1 \geq (n-1)p
\Leftrightarrow n-k-1 -pk \geq (n-1)p -pk\\
\Leftrightarrow n-k-1 -pk \geq p(n-k-1)
\Leftrightarrow 1-\frac{pk}{n-k-1} \geq p
\Leftrightarrow 1-p \geq \frac{pk}{n-k-1}.
$
Therefore, we have that, if $p\leq 1-\frac{k}{n-1}$, then $\lambda=\frac{pk}{n-k-1}$, else $\lambda=1-p$. In the latter, we have that $\epsilon=\sqrt{\frac{d\ln (n-2)}{(1-p)^2(n-2)}}$ and, therefore,\vspace{-0.4cm} $$\displaystyle \lim_{n\rightarrow \infty} \epsilon =\displaystyle \lim_{n\rightarrow \infty}\sqrt{\frac{d\ln (n-2)}{(1-p)^2(n-2)}}=0.$$

Now, let us consider the case $p\geq 1-\frac{k}{n-1}$. We have that $\lambda=\frac{pk}{n-k-1}$ and, therefore, $$\epsilon=\sqrt{\frac{d(n-k-1)^2\ln (n-2)}{p^2 k^2 (n-2)}}.$$ It is clear that, if the value of $k$ does not depend on $n$, the value of $\epsilon$ will diverge. So we need to analyze the behavior of $\epsilon$ when $k$ is a function of $n$.

Recall that $k$ is the number of initial neighbors in the $k$-connected ring lattice. Thus, $k/n$ represents the fraction of nodes in the network to which each node is initially connected. Let us consider the case of $\frac{k}{n}\geq \frac{1}{\ln^a(n)}, \forall n\geq n_0$ for some $a>0$ and $n_0\in \mathbb{N}$ (notice this includes the case of $\frac{k}{n}$ constant). In the following, all inequalities are considered to be for $n\geq n_0$. We have that\vspace{-0.5cm}
\begin{eqnarray}
\epsilon &=& \sqrt{\frac{d(n-k-1)^2\ln (n-2)}{p^2 k^2 (n-2)}} \leq \sqrt{\frac{d(n-2)^2\ln (n-2)}{p^2 k^2 (n-2)}}
= \sqrt{\frac{dn\ln (n-2)}{p^2 k^2}} \nonumber\\
&\stackrel{(a)}{\leq}& \sqrt{\frac{dn\ln (n-2)\cdot \ln^a(n)}{p^2 n^2}}
\leq \sqrt{\frac{d\ln^{b+1}(n)}{p^2 n}}, \label{eq:lim1}
\end{eqnarray}
where (a) follows from the fact that $k\geq \frac{n}{\ln^b(n)}$.

We have that $\displaystyle \lim_{n\rightarrow \infty}  \sqrt{d\ln^{b+1}(n)/p^2 n} =0$. Thus, using inequality \eqref{eq:lim1}, we have that $\displaystyle \lim_{n\rightarrow \infty} \epsilon =0$.

Now let us consider the case of $\frac{k}{n}\leq \frac{b}{n},\: \forall n\geq n_1$, for some $b>0$ and $n_1\in \mathbb{N}$ (notice that this is equivalent to $k\leq b$ and, therefore, includes the case where $k$ does not depend on $n$). In the following, all inequalities are considered to be for $n\geq n_1$. We have that
\begin{eqnarray}
\epsilon &=& \sqrt{\frac{d(n-k-1)^2\ln (n-2)}{p^2 k^2 (n-2)}}
\stackrel{(a)}{\geq} \sqrt{\frac{d(n-b-1)^2\ln (n-2)}{p^2 b^2 (n-2)}}, \label{eq:lim2}
\end{eqnarray}
where (a) follows from the fact that $k\leq b$.

We have that $\displaystyle \lim_{n \rightarrow \infty} \sqrt{\frac{d(n-b-1)^2\ln (n-2)}{p^2 b^2 (n-2)}} = \infty$. Therefore, using inequality \eqref{eq:lim2}, we have that $\displaystyle \lim_{n\rightarrow \infty} \epsilon =\infty$.
\end{proof}

\section{Dual Radio Networks}
\label{sec:drn}
This section is devoted to the probabilistic characterization of the max-flow min-cut capacity of
dual radio networks, which we model as follows.
\begin{definition}
\label{defi:drn}
A \textit{Dual Radio Network} (DRN) is a graph $G\left(
n,p,r_s,r_L\right)=\left( V,E\right) $ constructed by the following procedure.
Assign $n$ nodes uniformly at random in the set $T$, where $T$ is the torus
obtained by identifying the opposite sides of the box $\left[ 0,1\right] ^2$,
and define $V$ as the set of these $n$ nodes. For a parameter $r_S$, each pair
of nodes $(a,b),\textrm{ with }a,b\in V$, is connected if their euclidean
distance verifies $d\left(a,b\right)\leq r_S$, and let $E_S$ be the set of edges
created in this step. For a parameter $p$, define the set $V_L$ such that
$\forall i\in V$, $i\in V_L$ with probability $p$. For a parameter $r_L$, each pair of nodes
$(a,b),\:a,b\in V_L$ is connected if their Euclidean distance verifies
$d\left(a,b\right)\leq r_L$. Let $E_L$ be the set of edges created in this
step. Finally, the set of edges of a DRN is defined by $E= E_S \cup E_L$.
\end{definition}

\begin{figure}[p]
\begin{center}
\includegraphics[width=9cm]{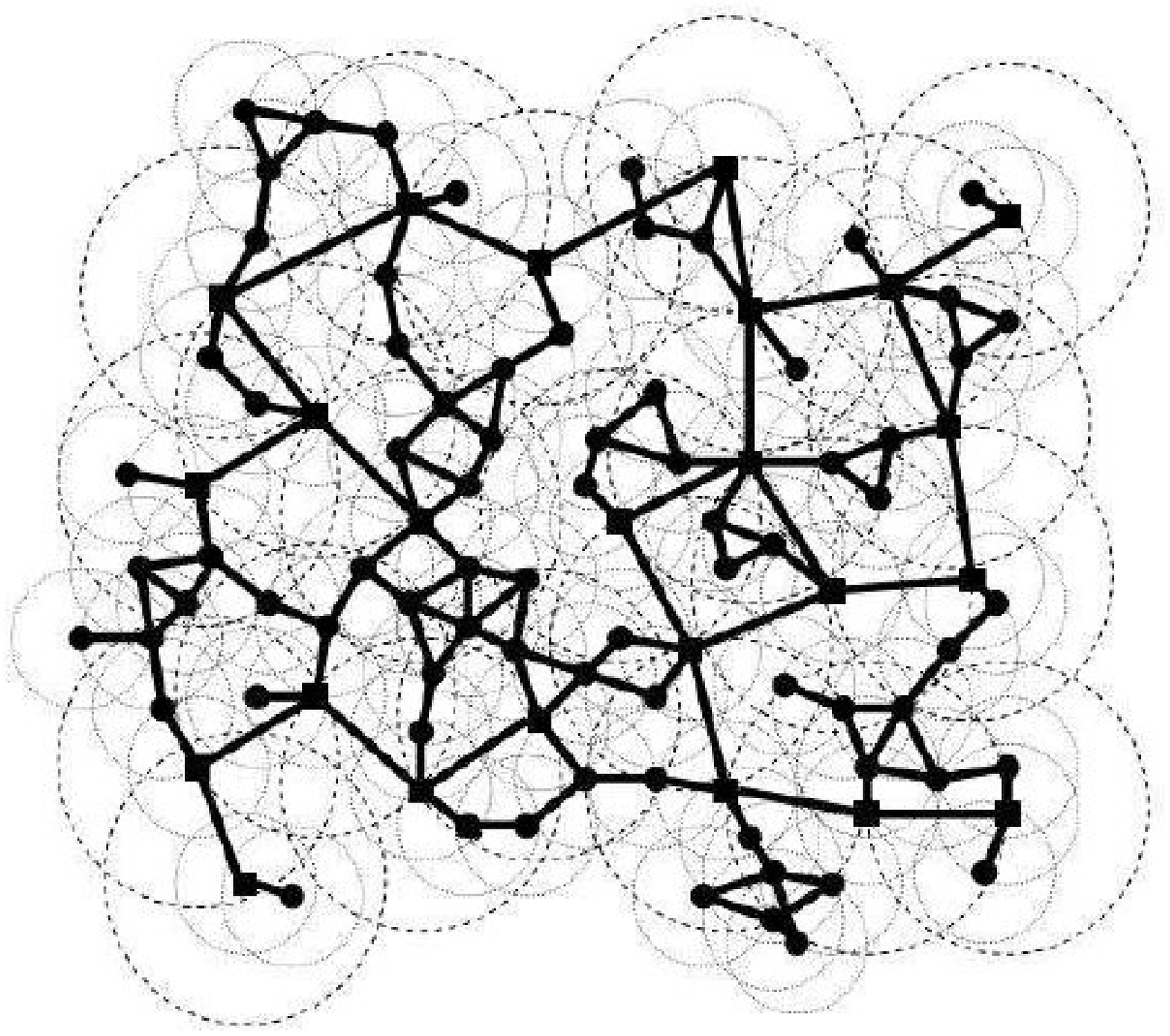}
\vspace{0.3cm}
\caption{Illustration of Dual Radio Networks. The square nodes represent the devices with two wireless technologies, whereas devices with only one
wireless technology are represented by circles. The small and large circumferences represent the coverage area of the short-range and of the long-range wireless interfaces, respectively.}
\label{fig:model}
\end{center}
\vspace{0.3cm}
\end{figure}

\fig{fig:model} provides an illustration of Dual Radio Networks. In the definition above, notice that any two nodes $a,b\in V$ satisfying $r_S <d(a,b)\leq r_L$ are connected if and only if both are elements of the set $V_L$. In light of the properties of the wireless networks that this graph model attempts to capture, this is a reasonable assumption since devices with a particular wireless technology can only establish links with other devices that possess a similar wireless interface.

The main result of this section is given by the following theorem.
\begin{theorem}
\label{th:drn}
The $s$-$T$-capacity of a Dual Radio Network, denoted by $C_{s;T}^{\textrm{DRN}}$, satisfies the following inequalities:\vspace{-0.2cm}
\begin{eqnarray*}
C_{s;T}^{\textrm{DRN}}&>&
(1-\epsilon)(n-2)\mu \textrm{ with probability }1-\textrm{O}\left(\frac{\alpha}{n^{2d}}
\right)\\
C_{s;T}^{\textrm{DRN}}&<&(1+\epsilon)(n-2)\mu \textrm{ with probability } 1-\textrm{O}\left(\frac{1}{n^{2d}} \right),
\end{eqnarray*}
for $\mu=\pi r_S^2+\pi p^2(r_L^2-r_S^2)$, $\epsilon=\sqrt{\frac{d\ln (n-2)}{\mu^2(n-2)}}$ and $1<d<\frac{\mu^2(n-2)}{\ln(n-2)}$. Moreover, $\displaystyle \lim_{n\rightarrow \infty} \epsilon =0$.
\end{theorem}

\begin{remark}
Clearly, for small enough values of $r_S$ and $r_L$, the network will be disconnected with high probability. Thus, its capacity is zero, which is captured in \theorref{th:drn}. If $r_S^2+p^2(r_L^2-r_S^2)<\sqrt{\frac{d\ln(n-2)}{\pi^2(n-2)}},$ then $\epsilon >1$ and, consequently, $(1-\epsilon)(n-2)\mu<0$, i.e.~ the bounds include zero.
\end{remark}

\begin{remark}
The bounds for the capacity of Dual Radio Networks are centered around $(n-2)\mu$, with $\mu=\pi r_S^2+\pi p^2(r_L^2-r_S^2)$, where $p$ represents the fraction of nodes with two different wireless interfaces. Thus, since $\displaystyle \lim_{n\rightarrow \infty} \epsilon =0$, we can say that in the limit of large networks the capacity of Dual Radio Networks grows quadratically with the number of nodes with two wireless interfaces.
\end{remark}

Before presenting the proof of \theorref{th:drn}, we need to state and prove some auxiliary results.
\begin{lemma}
\label{le:prob}
The probability of two nodes being connected in an instance of a Dual Radio Network is given by $\mu =\pi r_S^2+\pi p^2 (r_L^2-r_S^2)$.
\end{lemma}
\begin{proof}
First, we calculate the probability that a node ${\bf Y}$ is connected to node
${\bf X}$,
given the position of ${\bf X}$. This probability is given by\vspace{-0.3cm}
$$\prob ({\bf X}\leftrightarrow {\bf Y}|{\bf
X}) = \prob   \left(
\{d\left( {\bf X},{\bf Y}
\right) \leq  r_S\} \cup
 \left( \{ {\bf X} \in  V_L \}
 \cap  \{ {\bf Y} \in
 V_L\}  \cap  \{ d\left( {\bf
X},{\bf Y} \right) \leq r_L\} \right) \right|{\bf
X}).\vspace{-0.3cm}$$
Using the notation $\prob \left( A|{\bf X} \right)=\prob _{\bf X} (A)$
and $d({\bf X},{\bf Y})=D$, we have the
following:\vspace{-0.3cm}
\begin{eqnarray*}
\prob_{\bf X} ({\bf X} \leftrightarrow {\bf Y} )
&\stackrel{(a)}{=}& \prob_{\bf X}
\left(
D \leq r_S\right)+\prob_{\bf X}
\left(
\{ {\bf X}\in  V_L\} \cap
\{ {\bf Y}\in
V_L\}\cap
\{ D \leq r_L \}
\right)\\&&- \prob_{\bf X} \left( \{
D
\leq r_S \} \cap \{ {\bf X}
\in V_L \} \cap \{
{\bf Y} \in V_L \} \cap
\{ D  \leq  r_L \} \right)\\
&\stackrel{(b)}{=}& \prob_{\bf X}  \left(
D \leq r_S\right)
+\prob_{\bf X}  \left( \{ {\bf X}
\in
V_L\} \cap \{ {\bf
Y} \in
V_L\} \cap \{ D \leq
r_L \} \right)\\ && - \prob_{\bf X}
\left( \{ D \leq r_S \} \cap
 \{ {\bf X} \in V_L \}
\cap  \{ {\bf Y} \in  V_L \}
\right)
\end{eqnarray*}
where (a) follows from the fact that for any two events $A$ and $B$, $\prob
(A\cup B)=\prob(A)+\prob(B)-\prob(A\cap B)$, and (b) is justified by noting
that $D\leq r_S \Rightarrow D\leq r_L$, thus $\{D\leq r_S\}\cap \{D\leq
r_L\}=\{D\leq r_S \}$. The events $\{D\leq r_L \}$ and $\{ {\bf X}\in V_L\}$ are independent, and the
same is
true for the events $\{D\leq r_L \}$ and $\{ {\bf Y}\in V_L\}$. Because the set
of
nodes $V_L$ is formed by nodes selected at random and in an independent fashion,
we have that the events $\{ {\bf X}\in V_L\}$ and $\{ {\bf Y}\in V_L\}$ are
independent.
Therefore, $\prob_{\bf X}(\{{\bf X}\in V_L\}\cap\{{\bf
Y} \in V_L\} \cap \{ D \leq r_L
\})=\prob_{\bf X} ({\bf X} \in V_L) \cdot \prob_{\bf X}({\bf Y} \in V_L) \cdot \prob_{\bf X} (D \leq r_L).$ Using analogous arguments, we have that\vspace{-0.3cm}
\begin{eqnarray*}
\prob _{\bf X}(\{{\bf X}\in V_L\} \cap \{
{\bf Y}\in V_L\}\cap \{ D\leq r_S \} )=\prob_{\bf X} ({\bf X} \in V_L)
\cdot \prob_{\bf X}({\bf Y} \in V_L) \cdot \prob_{\bf X} (D\leq r_S).
\end{eqnarray*}

\vspace{-0.3cm}Noticing that the events $\{ {\bf X}\in V_L\}$ and $\{{\bf Y}\in V_L\}$ are
independent of the position of ${\bf X}$, we have that $\prob_{\bf X} ({\bf X}
\leftrightarrow {\bf Y} ) = \prob_{\bf X} (D\leq r_S)+\prob ({\bf X}\in V_L)
\cdot \prob ({\bf Y}\in V_L)\cdot \left( \prob_{\bf X} (D\leq r_L) - \prob_{\bf
X} (D\leq r_S)\right).$

Because the nodes are placed on a torus, we have that $\prob_{\bf
X} (D\leq \rho)=\pi \rho ^2$, with $\rho \leq 1/\sqrt{\pi}$. Noticing that
$\prob ({\bf X}\in V_L)=\prob ({\bf Y}\in V_L)=p$, we have that:\vspace{-0.3cm}
\begin{eqnarray}
\label{eq:probcone}
\prob_{\bf X}
({\bf X}\leftrightarrow {\bf Y} )= \pi r_S^2+\pi p^2 (r_L^2-r_S^2).
\end{eqnarray}

\vspace{-0.3cm}Let pos(${\bf X}$) be the random variable that represents the position of node ${\bf X}$. The final result follows from\vspace{-0.4cm}
\begin{eqnarray*}
\prob({\bf X}\leftrightarrow {\bf Y} )&=& \int_{[0,1]^2} \prob ({\bf X}\leftrightarrow {\bf Y} | \textrm{pos}({\bf X})=A) \cdot f_{\textrm{pos}({\bf X})}(A) dA \\
&=& \int_{[0,1]^2} (\pi r_S^2+\pi p^2 (r_L^2-r_S^2)) \cdot f_{\textrm{pos}({\bf X})}(A) dA \\
&=& (\pi r_S^2+\pi p^2 (r_L^2-r_S^2)) \cdot \int_{[0,1]^2} f_{\textrm{pos}({\bf X})}(A) dA
= \pi r_S^2+\pi p^2 (r_L^2-r_S^2).
\end{eqnarray*}

\vspace{-0.4cm}
\end{proof}

The next result shows that the Dual Radio Network model possesses the independence-in-cut property. It is a non-trivial example of a large class of networks whose edges are not independent random variables. Consider three nodes, ${\bf X}_1$, ${\bf X}_2$ and ${\bf X}_3$. Suppose that the position of ${\bf X}_1$ is known and that this node is connected both to ${\bf X}_2$ and ${\bf X}_3$. It follows that the positions of ${\bf X}_2$ and ${\bf X}_3$ are constrained by the position of ${\bf X}_1$, and thus  $\prob({\bf X}_2\leftrightarrow {\bf X}_3|{\bf X}_1\leftrightarrow {\bf X}_2,{\bf X}_1\leftrightarrow {\bf X}_3,{\bf X}_1={\bf x})\neq \prob({\bf X}_2\leftrightarrow {\bf X}_3|{\bf X}_1={\bf x})$. However, as the next result shows, the edges across a given cut in a Dual Radio Network are independent random variables.

\begin{lemma}
\label{le:inde}
A Dual Radio Network exhibits the independence-in-cut property.
\end{lemma}

\begin{proof}
We will start by showing that the outgoing edges of a node ${\bf X}$ are independent random variables, when conditioned on the position of ${\bf X}$. This means that $\{{\bf X}\leftrightarrow {\bf
Y}_1\},\{{\bf X}\leftrightarrow {\bf Y}_2\},\dots,\{{\bf X}\leftrightarrow {\bf Y}_{n-1}\}$ are mutually independent conditioned on the fact that the position of node ${\bf X}$ is fixed i.e.~ ${\bf X}={\bf x}$). Without loss
of generality, we may write:\vspace{-0.4cm}
$$\prob ({\bf
X}\leftrightarrow {\bf Y}_1|{\bf X}\leftrightarrow {\bf Y}_2,\dots,{\bf
X}\leftrightarrow {\bf Y}_{n-1},{\bf X}={\bf x}) =\prob ({\bf Y}_1\leftrightarrow {\bf x}|{\bf Y}_2\leftrightarrow
{\bf x},\dots,{\bf Y}_{n-1}\leftrightarrow {\bf x}),\vspace{-0.4cm}
$$
where we exploited the fact that the position of ${\bf X}$ is fixed.
Now, notice that none of the events $\{ {\bf Y}_2\leftrightarrow {\bf x}\},\dots,\{ {\bf
Y}_{n-1}\leftrightarrow {\bf x}\}$ affects the event $\{ {\bf Y}_1\leftrightarrow
{\bf x}\}$, because we do not have information about the existence of connection
between ${\bf Y}_1$ and any of the ${\bf Y}_i$. Therefore, $\prob ({\bf
X}\leftrightarrow {\bf Y}_1|{\bf X}\leftrightarrow {\bf Y}_2,\dots,{\bf
X}\leftrightarrow {\bf Y}_{n-1},{\bf X}={\bf x})=\prob ({\bf X}\leftrightarrow
{\bf Y}_1|{\bf X}={\bf x}).$ Since we can use similar arguments for
different subsets of the collection $\left\{ \{{\bf X}\leftrightarrow {\bf
Y}_1\},\{{\bf X}\leftrightarrow {\bf Y}_2\},\dots,\{{\bf X}\leftrightarrow {\bf
Y}_{n-1}\}\right\}$, we have that the events $\{{\bf X}\leftrightarrow {\bf
Y}_1\},\{{\bf X}\leftrightarrow {\bf Y}_2\},\dots,\{{\bf X}\leftrightarrow {\bf
Y}_{n-1}\}$ are mutually independent, conditioned on the fact that the position of node ${\bf X}$ is fixed.
Consider a $s$-$t$-cut of size $k$, $C_k$. Consider a set of nodes $\{{\bf X},{\bf Y}_1,\dots,{\bf Y}_m\}$. We have that\footnote{Similar arguments are used in~\cite{ramamoorthy:05}.}\vspace{-0.3cm}
\begin{eqnarray}
\prob (C_{{\bf X}{\bf Y}_1}=&z_1&,C_{{\bf X}{\bf Y}_2}=z_2,\dots,C_{{\bf X}{\bf Y}_m}=z_m) \nonumber \\&=& \int_{[0,1]^2}\prob(C_{{\bf X}{\bf Y}_1}=z_1,C_{{\bf X}{\bf Y}_2}=z_2,\dots,C_{{\bf X}{\bf Y}_m}=z_m |\textrm{pos}({\bf X})=A)\cdot f_{\textrm{pos}({\bf X})}dA\nonumber \\
&\stackrel{(a)}{=}& \int_{[0,1]^2} \prod_{\alpha=1}^m \prob(C_{{\bf X}{\bf Y}_{\alpha}}=z_{\alpha}|\textrm{pos}({\bf X})=A) \cdot f_{\textrm{pos}({\bf X})}dA\nonumber \\
&\stackrel{(b)}{=}& \int_{[0,1]^2} \prod_{\alpha=1}^m \mu^{z_{\alpha}}(1-\mu)^{1-z_{\alpha}} \cdot f_{\textrm{pos}({\bf X})}dA\nonumber \\
&\stackrel{(c)}{=}& \prod_{\alpha=1}^m \mu^{z_{\alpha}}(1-\mu)^{1-z_{\alpha}} \cdot \int_{[0,1]^2} f_{\textrm{pos}({\bf X})}dA\label{eq:posinde} \\
&=& \prod_{\alpha=1}^m \mu^{z_{\alpha}}(1-\mu)^{1-z_{\alpha}}=\prod_{\alpha=1}^m \prob (C_{{\bf X}{\bf Y}_{\alpha}}=z_{\alpha}),\nonumber
\end{eqnarray}
where we used the following arguments:
\begin{itemize}
 \item (a) follows from the fact that outgoing edges of a node are independent, as we have already demonstrated;
 \item (b) follows from the property that two nodes are connected with probability is $\mu$, thus
\begin{displaymath}
\prob(C_{{\bf X}{\bf Y}_{\alpha}}=z_{\alpha}|\textrm{pos}({\bf X})=A)=\mu^{z_{\alpha}}(1-\mu)^{1-z_{\alpha}}=\left\{ \begin{array}{ll}
\mu & \textrm{if }z_{\alpha}=1\\
1-\mu & \textrm{if }z_{\alpha}=0\\
\end{array}\right. ;
\end{displaymath}
 \item (c) follows from the fact that $\mu^{z_{\alpha}}(1-\mu)^{1-z_{\alpha}}$ does not depend on the position of ${\bf X}$.
\end{itemize}

This reasoning shows that the outgoing edges of any given node are independent. Using similar arguments, we can also prove that the incoming edges of any given node are also independent. Recall that a cut $C_k$ is of the form $C_k=\sum\limits_{i\in \overline{V}_k}C_{si}+\sum\limits_{j\in V_k}\sum\limits_{i\in \overline{V}_k}C_{ji}+\sum\limits_{j\in V_k}C_{jt}$. With independent outgoing edges and incoming edges for any given node and knowing that edges with no node in common are also independent, we have that all the edges that cross the cut are independent random variables, which clearly satisfies the definition of the independence-in-cut property.
\end{proof}

We are now ready to prove \theorref{th:drn}.

\begin{proofdrn}
We start by calculating the minimum expected value of a cut in an instance of a Dual Radio Network. Consider a $s$-$t$-cut of size $k$, $C_k=\sum\limits_{i\in \overline{V}_k}C_{si}+\sum\limits_{j\in V_k}\sum\limits_{i\in \overline{V}_k}C_{ji}+\sum\limits_{j\in V_k}C_{jt}$. Thus, we have that $E(C_k)=\sum\limits_{i\in \overline{V}_k}E(C_{si})+\sum\limits_{j\in V_k}\sum\limits_{i\in \overline{V}_k}E(C_{ji})+\sum\limits_{j\in V_k}E(C_{jt})$. Hence, because $E(C_{ij})=\prob(i\leftrightarrow j)=\mu,\: \forall i,j$ by \lemmaref{le:prob}, we have that $E(C_k)=(N+x(N-x))\mu$. Therefore, we have that $c_{\min}=\displaystyle \min_{k\in \{0,\dots,N \}} (N+x(N-x))\mu=N\mu$, which yields $c_{\min}=(n-2)\mu.$
Now, notice that $\lambda= \displaystyle \min_{i,j:\prob(i\leftrightarrow j)>0}{\prob(i\leftrightarrow j)}=\mu$. Thus, since we have proven that a Dual Radio Network has the independence-in-cut property in \lemmaref{le:inde}, we are ready to use~\theorref{th:main} and the bounds follow.

To conclude the proof of the theorem, just notice that $\displaystyle \lim_{n \rightarrow \infty}\epsilon=\displaystyle \lim_{n \rightarrow \infty} \sqrt{\dfrac{d\ln (n-2)}{\mu^2(n-2)}}=0$.
\vspace{-0.4cm}
\end{proofdrn}

\proofend

The previous result presents bounds for the capacity of a Dual Radio Network, where it was assumed (see \definref{defi:drn}) that the metric used was a wrap-around metric in a unit square, i.e. the space considered was a torus, which is obtained by identifying the opposite sides of the box $\left[ 0,1\right] ^2$. In particular, it was assumed that all the nodes in the network have the same area of coverage. If we do not consider a torus but instead the standard $\left[ 0,1\right] ^2$ square, it is clear that nodes close to the border will have a smaller area of coverage. In this case, a Dual Radio Network will no longer have the independence-in-cut property. In fact, in the proof of \lemmaref{le:inde}, namely in equality \eqref{eq:posinde}, it is crucial that the probability of two nodes being connected is independent of the position of the nodes, which means that all nodes have to have the same coverage area. The fact that, in the wrap-around case, Dual Radio Networks have the independence-in-cut property was crucial for obtaining the bounds for the capacity of these networks. In the following, we will provide bounds on the capacity of Dual Radio Networks based on a non-wrap-around square by analyzing networks that are similar to Dual Radio Networks, but where all the nodes obtain the same coverage area, by increasing (or decreasing) their radio range.

\begin{theorem}
Consider a Dual Radio Network generated in the unit square $[0,1]^2$ with a Euclidean metric (i.e. with no wrap-around). The $s$-$T$-capacity of this network, denoted by $C_{s;T}^{\textrm{DRN}^*}$, satisfies the following inequalities
\begin{eqnarray*}
C_{s;T}^{\textrm{DRN}^*}&>&
(1-4\epsilon)(n-2)\frac{\mu}{4} \textrm{ with probability }1-\textrm{O}\left(\frac{\alpha}{n^{2d}}
\right)\\
C_{s;T}^{\textrm{DRN}^*}&<&(1+\epsilon)(n-2)\mu \textrm{ with probability } 1-\textrm{O}\left(\frac{1}{n^{2d}} \right),
\end{eqnarray*}
for $u=\pi r_S^2+\pi p^2(r_L^2-r_S^2)$, $\epsilon=\sqrt{\frac{d\ln (n-2)}{\mu^2(n-2)}}$ and $1<d<\frac{\mu^2(n-2)}{\ln(n-2)}$. Moreover, $\displaystyle \lim_{n\rightarrow \infty} \epsilon =0$.
\end{theorem}

\begin{proof}
The main idea of the proof is to consider the situation of nodes adjusting their transmitting range so that all the nodes have the same coverage area\footnote{This technique is also used in \cite{ramamoorthy:05} to prove results for Random Geometric Graphs.}. In a Dual Radio Network based on the unit square with an Euclidean metric, nodes closer to the border have lower coverage area than nodes in the center of the square. More precisely, we have that the corner nodes have the minimum coverage area from all the nodes. Using arguments similar to those used to prove \lemmaref{le:prob}, namely Equality \eqref{eq:probcone}, we have that $\prob ({\bf X}\leftrightarrow{\bf Y}|{\bf X}\textrm{ is a corner node})= \frac{\pi}{4} r_S^2+\frac{\pi}{4} p^2(r_L^2-r_S^2).$

Consider the situation where all nodes adjust their communication range such that $\forall {\bf X},\: \prob ({\bf X}\leftrightarrow{\bf Y}|{\bf X}={\bf x})=\frac{\pi}{4} r_S^2+\frac{\pi}{4} p^2(r_L^2-r_S^2)=\mu '$ and let $C_{s;T}'$ be the $s$-$T$-capacity of this network. This means that all nodes (except the corner ones) have to reduce their transmitting power (for both wireless communication technologies). In this case, we have that the probability of two nodes being connected does not depend on the position, thus the proof of \lemmaref{le:inde} holds and, therefore, this network has the independence-in-cut property. Moreover, if $C_k$ is a cut of size $k$, we have that $E(C_k)= (N+x(N-x))\mu'$, since $E(C_{ij})=\prob(i\leftrightarrow j)=\mu',\:i,j$. Therefore, $c_{\min}=(n-2)\mu'$ and, by \theorref{th:main}, $\prob(C_{s;T}'\leq (1-\epsilon')(n-2)\mu')= \textrm{O}\left(\frac{\alpha}{n^{2d}}
\right),$ with $\epsilon'=\sqrt{\frac{d\ln (n-2)}{\mu'^2(n-2)}}$. Notice that, with $\mu=\pi r_S^2+\pi p^2(r_L^2-r_S^2)$ and $\epsilon=\sqrt{\frac{d\ln (n-2)}{\mu^2(n-2)}}$, we have that $\mu'=\frac{\mu}{4}$ and, thus, $\epsilon'=4\epsilon$. Thus, we have that\vspace{-0.6cm}
\begin{eqnarray}
\label{eq:nsei21}
\prob\left(C_{s;T}'\leq (1-4\epsilon)(n-2)\frac{\mu}{4}\right)= \textrm{O}\left(\frac{\alpha}{n^{2d}}
\right).
\end{eqnarray}

In a Dual Radio Network, it is clear that many nodes will have an higher coverage area, which can only lead to an increase of capacity. Thus, we have that $C_{s;T}'\leq C_{s;T}^{\textrm{DRN}^*}.$ Therefore, if $C_{s;T}^{\textrm{DRN}^*}\leq (1-4\epsilon)(n-2)\frac{\mu}{4}$, then $C_{s;T}'\leq (1-4\epsilon)(n-2)\frac{\mu}{4}$, which implies that \vspace{-0.2cm}$$\prob\left(C_{s;T}^{\textrm{DRN}^*}\leq (1-4\epsilon)(n-2)\frac{\mu}{4} \right) \leq \prob\left(C_{s;T}'\leq (1-4\epsilon)(n-2)\frac{\mu}{4} \right).$$ Thus, from \eqref{eq:nsei21}, we have that $\prob\left(C_{s;T}^{\textrm{DRN}^*}\leq (1-4\epsilon)(n-2)\frac{\mu}{4} \right) =\textrm{O}\left(\frac{\alpha}{n^{2d}}
\right).$

Now, to compute the upper bound on $\prob \left(C_{s;T}^{\textrm{DRN}^*}\geq(1+\epsilon)(n-2)\mu\right)$, we will use the same approach as before, but now with some increasing their transmitting power. Consider the situation where all the nodes adjust their communication range such that $\forall {\bf X},\: \prob ({\bf X}\leftrightarrow{\bf Y}|{\bf X}={\bf x})=\frac{\pi}{4} r_S^2+\frac{\pi}{4} p^2(r_L^2-r_S^2)=\mu$ and let $C_{s;T}''$ be the $s$-$T$-capacity of this network. This means that the nodes closer to the border of the unit square will increase their transmitting power (for both technologies). As before, in this situation we have that the network has the independence-in-cut property and, with $c_{\min}=(n-2)\mu$, we can use \theorref{th:main} and we get\vspace{-0.1cm}
\begin{eqnarray}
\label{eq:nsei44}
\prob\left(C_{s;T}''\geq (1+\epsilon)(n-2)\mu\right)= \textrm{O}\left(\frac{\alpha}{n^{2d}}
\right)
\end{eqnarray}
with $\epsilon=\sqrt{\frac{d\ln (n-2)}{\mu^2(n-2)}}$. In the real situation, many nodes will have a lower transmitting power, which can only lead to a decrease in the capacity. Therefore, $C_{s;T}''\geq C_{s;T}^{\textrm{DRN}^*}.$ Therefore, if $C_{s;T}^{\textrm{DRN}^*}\geq (1+\epsilon)(n-2)\mu$, then $C_{s;T}''\geq (1+\epsilon)(n-2)\mu$, which implies that $\prob\left(C_{s;T}^{\textrm{DRN}^*}\geq (1+\epsilon)(n-2)\mu \right) \leq \prob\left(C_{s;T}''\geq (1+\epsilon)(n-2)\mu \right).$ Thus, by \eqref{eq:nsei44}, we have that $\prob\left(C_{s;T}^{\textrm{DRN}^*}\geq (1+\epsilon)(n-2)\mu \right) =\textrm{O}\left(\frac{1}{n^{2d}}
\right).$
To conclude the proof of the theorem, just notice that $\displaystyle \lim_{n \rightarrow \infty}\epsilon=\displaystyle \lim_{n \rightarrow \infty} \sqrt{\dfrac{d\ln (n-2)}{\mu^2(n-2)}}=0$.
\end{proof}

\section{Conclusions}
\label{sec:conclusions}
After defining the property of {\it independence-in-cut} for the edges of random graphs,
we proved a general theorem that provides upper and lower bounds for the max-flow min-cut
capacity of any graph satisfying this property. This theorem is not only of some interest by
itself, but also proves to be a valuable tool in determining capacity bounds for small-world graphs
and dual radio models, whose importance stems from their arguably compelling applications.
Perhaps the most striking conclusions to be drawn from our results are that (a) rewiring
satisfying the independence-in-cut property does not change the capacity of a small-world network up to a constant factor and that
(b) the capacity of dual radio networks grows quadratically with the fraction of dual radio devices,
thus indicating that a small percentage of such devices is sufficient to improve significantly
the maximum information flow in the network.
\bibliographystyle{IEEE}
\bibliography{pp}

\end{document}